\documentclass[journal]{IEEEtran}
\newif\ifReviewerview 
\Reviewerviewfalse

\newcommand{\FigWidth}{1.0}

\usepackage{graphicx}
\usepackage{dcolumn}
\usepackage{bm}
\usepackage{amsmath,amsfonts,amssymb}

\usepackage[version=4]{mhchem}

\usepackage[product-units=bracket-power,
            range-units  =single,
            range-phrase =--]
            {siunitx}
\usepackage{cite}
\usepackage{makecell}
\usepackage[switch,columnwise]{lineno}

\usepackage[T1]{fontenc}
\usepackage[hidelinks]{hyperref}
\usepackage{orcidlink}
\usepackage{mathptmx}
\usepackage{etoolbox}
\usepackage{mhchem}
\usepackage[table]{xcolor}
\usepackage[normalem]{ulem}
\usepackage{booktabs} 
\usepackage{tabularx}
\usepackage{array}
\usepackage{mathtools}
\usepackage[
math-style=ISO,
warnings-off={mathtools-colon, mathtools-overbracket},
]{unicode-math}
\usepackage{mathptmx} 

\usepackage{footnote} 
\usepackage{csquotes}

\ifReviewerview
    \newcommand{\rem}[1]{\textcolor{red}{V0: #1}} 
    \newcommand{\JH}[1]{\textcolor{blue}{Rev1: #1}}
    \linenumbers
\else
        \newcommand{\rem}[1]{\textcolor{black}{}} 

        \NewDocumentCommand{\JH}{+m}{%
          \begingroup
          \color{black}%
          #1%
          \endgroup
        }
\fi 

\newcolumntype{M}[1]{>{\centering\arraybackslash}m{#1}}

\NewDocumentCommand{\nep}{mo}{%
  \qty{#1}{\pico\watt\per\sqrt{Hz}}%
}
\NewDocumentCommand{\nepuW}{mo}{%
  \qty{#1}{\micro\watt\per\sqrt{Hz}}%
}
\NewDocumentCommand{\nepnW}{mo}{%
  \qty{#1}{\nano\watt\per\sqrt{Hz}}%
}
\NewDocumentCommand{\nepmW}{mo}{%
  \qty{#1}{\milli\watt\per\sqrt{Hz}}%
}
\NewDocumentCommand{\nVperHz}{mo}{%
  \qty{#1}{\nano\volt\per\sqrt{Hz}}%
}

\NewDocumentCommand{\neprange}{mm}{%
  \qtyrange{#1}{#2}{\pico\watt\per\sqrt{Hz}}%
}

\usepackage[nameinlink,capitalize]{cleveref}
\makeatletter
\begingroup
\catcode`\_=\active
\protected\gdef_{\@ifnextchar|\subtextit\subtextup}
\endgroup
\def\subtextit|#1|{\sb{#1}}
\def\subtextup#1{\sb{\mathrm{#1}}}
\AtBeginDocument{\catcode`\_=12 \mathcode`\_=32768}
\makeatother

\begin{document}

\title{Improving terahertz-detection sensitivity of \(8 \times 8\)~FET arrays through liquid-nitrogen cooling in a compact low-noise cryostat}
\author{Jakob~Holstein~\orcidlink{0009-0001-8744-475X},
Nicholas~K.~North~\orcidlink{0009-0002-8221-2893},
Arne~Hof~\orcidlink{0009-0001-8046-7377},
Sanchit~Kondawar~\orcidlink{0000-0003-3918-9472}, 
Dmytro~B.~But~\orcidlink{0000-0002-0735-4608},
Mohammed~Salih~\orcidlink{0009-0001-6882-4642},
Lianhe~Li~\orcidlink{0000-0003-4998-7259},
Edmund~H.~Linfield~\orcidlink{0000-0001-6912-0535},
A.~Giles~Davies~\orcidlink{0000-0002-1987-4846},
Joshua~R.~Freeman~\orcidlink{0000-0002-5493-6352},
Alexander~Valavanis~\orcidlink{0000-0001-5565-0463},
Alvydas~Lisauskas~\orcidlink{0000-0002-1610-4221
}, \IEEEmembership{Member, IEEE},
and
Hartmut~G.~Roskos~\orcidlink{0000-0003-3980-0964}
\thanks{
Manuscript received 18~July, 2025; revised 9~Jan, 2026; accepted 06~Feb, 2026. Date of publication ??~????, 2026; date of current version 06~Feb, 2026.
This work was supported by the German Research Foundation (DFG) project RO 770 49-1 of the INTEREST collaborative research program program (SPP~2314), UKRI (Future Leader Fellowship MR/Y011775/1 and MR/S016929/1) and EPSRC (programme grant “TeraCom”, EP/W028921/1). 
AL acknowledges funding received from the Lithuanian Science Foundation (project No. S-MIP-22-83).
DB acknowledges funding by European Union (ERC ”TERAPLASM”, project number: 101053716) and the European Regional Development Fund (the PolFEL project No. KPOD.01.18-IW.03-0019/23).
For the purpose of open access, the author has applied a Creative Commons Attribution (CC BY) license to any Author Accepted Manuscript version arising from this submission.} 

\thanks{J. Holstein, A. Hof and H. G. Roskos are with the Physikalisches Institut, Johann Wolfgang Goethe-Universität, DE-60438 Frankfurt am Main, Germany (e-mail: roskos@physik.uni-frankfurt.de)}
\thanks{A. Lisauskas is with (i) the Institute of Applied Electrodynamics and Telecommunications, Vilnius University, LT-10257 Vilnius, Lithuania, (ii) Center for Physical Sciences and Technology, LT-10257 Vilnius, Lithuania, (iii) Physikalisches Institut, Johann Wolfgang Goethe-Universität, DE-60438 Frankfurt am Main, Germany.}
\thanks{N.~K.~North,
S.~Kondawar,
M.~Salih,
L.~Li,
E.~H.~Linfield,
A.~G.~Davies,
J.~R.~Freeman,
and A.~Valavanis are with the School of Electronic and Electrical Engineering, University of Leeds, Leeds LS2~9JT, United Kingdom. (e-mail: a.valavanis@leeds.ac.uk)}
\thanks{D. B. But is with (i) CENTERA, Institute of High Pressure Physics PAS,  01-142 Warsaw, Poland, (ii) IB1, National Centre for Nuclear Research Świerk, 05-400 Otwock, Poland}%
}

\date{\today}
\maketitle

\begin{abstract}
We show that the sensitivity of antenna-coupled field-effect transistors (FETs) to terahertz (THz) radiation improves continuously with decreasing temperature.
The noise-equivalent power (NEP) of \qty{540}{\GHz} patch-antenna-coupled FETs decreases as temperature reduces to \qty{20}{K}.
We project NEP values approaching \qtyrange{1}{2}{\pico\watt\per\sqrt\Hz} under efficient power coupling conditions \JH{(e.g., using a superstrate Si-lens)}, which is comparable to superconducting niobium transition-edge sensors (TESs) at \qty{4}{K}.
Building on these findings, a compact, low-noise, liquid-nitrogen-cooled (\qty{77}{K}) \JH{FET-based direct (incoherent) THz-power sensing system} for spectroscopy applications was realized. 
Here, an 8\(\times\)8 pixel-binned detector array fabricated in a commercial 65\mbox{-}nm Si-CMOS process, was optimized for operation in the \qtyrange{2.85}{3.4}{\THz} band.
Characterization was performed in the focal plane of a 2.85\mbox{-}THz quantum-cascade laser delivering \(\sim\)2~mW of THz power.
A linear dynamic range exceeding \qty{67}{dB} was achieved without saturation (for 1~Hz-detection bandwidth).
The system provides a \qty{-3}{dB} readout bandwidth of \qty{5}{MHz}, exceeding that of conventional thermal detectors (typically \qty{1}{kHz}).
Combined with its broad temperature operability \JH{(\qtyrange{20}{300}{\kelvin})} and compact design, the system is particularly well suited for space- and payload-constrained platforms such as balloon- and satellite-based missions, where deep cryogenic cooling is impractical.
\end{abstract}
\begin{IEEEkeywords}
terahertz, detection, MOSFET, Gaussian beam, power coupling, quantum-cascade lasers, gas spectroscopy, cryogenic systems, carrier freeze-out effect.
\end{IEEEkeywords}

\IEEEpeerreviewmaketitle

\section{Introduction}

\IEEEPARstart{T}{he} terahertz (THz) frequency range of the electromagnetic spectrum (\qtyrange{0.3}{10}{\THz}) has attracted growing attention over the past two decades, with potential applications including high-speed wireless communication, non-destructive testing, imaging, chemical or biological sensing~\cite{leitenstorfer_2023_2023}, 
and gas spectroscopy~\cite{hubers_high-resolution_2019,waters_observing_2024}.
However, the \(\sim\)\qtyrange{1}{5}{\THz} band remains technologically challenging owing to the limited availability of powerful THz sources and sensitive detectors or mixers that can operate without liquid-helium cooling.
This is particularly important for balloon-borne~\cite{wienold_osas-b_2024} and satellite-borne~\cite{savini_recent_2016} instrumentation, where the cooling infrastructure is constrained by strict limitations on payload mass, volume, and power consumption.
\JH{Cryogenically-cooled bolometers are extensively utilized for terahertz sensing applications, owing to their high sensitivity (low Noise Equivalent Power, NEP) (see the overview in \autoref{tab:addlabel}).
However, the main limitation of thermal detectors is their comparably slow detection mechanism~\cite{wubs_terahertz_2023}.
As such, the absence of fast and sensitive detectors---especially those capable of continuous, cryogen-free operation---remains a significant barrier to the broader application and integration of THz systems within this frequency range.}
In this context, we present a compact, low-noise, liquid-nitrogen (LN\(_2\))-cooled power detection system with a \qty{-3}{\dB} bandwidth of approximately \qty{5}{\MHz} (mainly limited by the chosen readout electronics), capable of detecting THz radiation emitted by a quantum-cascade laser (QCL) at \qty{2.85}{\THz}.
This is enabled by the core element of the system: an \(8 \times 8\) antenna-coupled field-effect transistor (FET) array employing a pixel binning technique, as reported in~\cite{holstein_88_2024}.
Over the past decades, THz detection using FETs has evolved from fundamental research devices~\cite{knap_nonresonant_2002,knap_field_2009,lisauskas_rational_2009} into high-performance detectors covering the entire terahertz (THz) frequency range~\cite{boppel_cmos_2012,regensburger_broadband_2018,yadav_gaas-based_2024}. 
Applications range from imaging~\cite{yuan_3d_2019,boppel_cmos_2012-1,lisauskas_exploration_2014,valusis_roadmap_2021}, spectroscopy~\cite{bauer_antenna-coupled_2014,krysl_si_2024,holstein_88_2024,horbury_real-time_2023}, tracking of ultrashort pulses~\cite{preu_ultra-fast_2013,lisauskas_field-effect_2018} or passive detection of thermal radiation~\cite{cibiraite-lukenskiene_passive_2020,andree_towards_2024}.
FETs can be employed either as direct THz power detectors or as heterodyne mixers~\cite{glaab_terahertz_2010,grzyb_thz_2015}, operating effectively in both room temperature and cryogenic environments~\cite{klimenko_temperature_2012,ikamas_optical_2020}.
Owing to their electronic detection mechanism---based on the excitation of plasma waves in the two-dimensional electron gas (2DEG) of the transistor channel---FETs exhibit an intrinsically high detection speed, with response times reported down to \qty{12}{ps}~\cite{lisauskas_field-effect_2018} corresponding to 10s of GHz achievable modulation bandwidth.

A key advantage over established \JH{electronic} THz detector technologies, such as commercially available Schottky-diode technology, is their compatibility with widely used semiconductor platforms including AlGaN/GaN~\cite{bauer_high-sensitivity_2019}, AlGaAs/GaAs~\cite{regensburger_broadband_2018}, or Si-CMOS~\cite{lisauskas_rational_2009,zdanevicius_field-effect_2018,ikamas_all-electronic_2021,ikamas_broadband_2018,ikamas_efficient_2017,javadi_sensitivity_2021,cesiul_towards_2022,holstein_88_2024}.
Graphene-based FETs are also a currently-emerging technology~\cite{generalov_400-ghz_2017,caridad_room-temperature_2024,ludwig_terahertz_2024,holstein_towards_2024}. 
This technological flexibility enables, for example, the fabrication of Si-CMOS-based FETs using mature commercial foundry processes, offering high scalability and process reliability.
In the detectors presented in this work, the 65-nm CMOS technology node provided by Taiwan Semiconductor Manufacturing Company (TSMC) was utilized.
This allows not only for the realization of individual detectors, but also for the cost-efficient production of large-scale detector arrays as e.g., required for THz camera systems~\cite{ojefors_065_2009,al_hadi_1_2012,liu_exploration_2023,holstein_88_2024}. 
These advantages make \JH{FET-based terahertz detectors} a promising candidate for future scientific and industrial applications.

This work is structured in the following way. 
Before presenting a detailed analysis of LN\(_2\)-cooled FETs for THz-QCL applications, we highlight the potential of cryogenic cooling through a systematic study that quantitatively investigates the NEP of a Si-CMOS FET coupled to a patch antenna resonant at \qty{540}{\GHz}.

The optical NEP of this detector at room temperature, when coupled to a superstrate Si-lens, can be as low as \nep{16}~\cite{krysl_si_2024}.
In \JH{the presented experiment}, no Si-lens was applied.
However, based on the experimental beam conditions, cross-sectional NEP on the order of \(~\sim\)\nep{25} was determined at room temperature for the  \SI{540}{\giga \hertz}-detector, which in is in good agreement with our previously published work~\cite{krysl_si_2024} [Table 1: Case A (without lens)].
We observe a continuous NEP-improvement down to \qty{20}{K}. Specifically, we report a fourfold improvement at \qty{77}{K}, which is in close agreement with previously published study investigating TeraFETs fabricated in \SI{90}{\nano \meter} Si-CMOS technology. 
Based on our results obtained with the TeraFET detector resonant at \SI{540}{\giga\hertz} and operated at \SI{20}{\kelvin}, we observed up to a fifteenfold improvement in NEP compared to room-temperature operation, corresponding to an expected NEP in the range of \neprange{1}{2}, assuming efficient power coupling ---  for instance, when using a superstrate silicon lens coupled to the patch antenna. This performance is comparable to that of state-of-the-art superconducting transition-edge sensor (TES) bolometers operated at \SI{4}{\kelvin}.
In \autoref{tab:addlabel}, we finally summarize the most important detector parameters investigated here and compare with other state-of-the-art power detector technologies available for the frequency range around \SI{3}{\tera \hertz}.

\section{FET-based THz detectors}
Current TeraFET detectors reach optical NEPs\footnote{Optical NEP refers to full available power in the experimental beam focus. It is a rather conservative measure of detector sensitivity, but is a good predictor for available sensitivity in real-world applications. An overview and critical discussion of different NEP calculations applied throughout the literature is given in~\cite{ikamas_all-electronic_2021,javadi_sensitivity_2021}.}
down to \neprange{10}{20} in the frequency range up to \qty{1}{\THz} at room temperature~\cite{jain_terahertz_2018,krysl_si_2024,ferreras_broadband_2021,wiecha_antenna-coupled_2021} --- a sensitivity level suitable for many practical applications. 
For applications in the frequency range above \(\sim\)\qty{2}{\tera \hertz}, where THz QCL sources can be employed, the sensitivity of available electronic detector (Schottky and \JH{FET}) technologies decreases.  However, while Schottky-based detectors experience strong frequency related roll-off due to large junction capacitance, FET sensitivity is expected to show less pronounced roll-off behavior~\cite{yadav_state---art_2023,yadav_gaas-based_2024}.

The effect of cooling TeraFET detectors to cryogenic temperatures is twofold.
First, the signal magnitude increases due to a more efficient mixing process~\cite{klimenko_temperature_2012}, which is driven by the enhanced non-linearity of the gated channel’s carrier density \(n(V_{GS})\) around the transistor’s threshold voltage.
\JH{Second, thermal noise from the FET's Ohmic resistance is reduced, according to \autoref{eq:JohnsonNyquistNoise}.
Mathematically, \(n(V_{GS})\) can be described with the Unified Charge Control Model (UCCM)~\cite{shur_unified_1992}.}
In our implementation, electromagnetic THz radiation is coupled to the transistor between source and gate electrode\footnote{Note that the gate to channel potential is a superposition of applied DC-bias \(V_{GS}\) and \(V_{THz}\cdot \cos{\omega_{THz}t}\), where $V_{THz}$ is the THz-radiation's amplitude being fed to the transistor from the patch-antenna element.} (see \autoref{fig:TeraFETcrossSectionB6}). This coupling scheme excites strongly damped carrier-density modulations (plasma waves) that propagate from the source into the gated channel.

\if{0}
The nonlinear mixing process of these plasma waves within the channel is strongly influenced by the inverse gradient of the logarithmic carrier density in the gated channel region,
\begin{equation}
\left(\frac{\partial \ln{[n(V_{GS})]}}{\partial V_{GS}}\right)^{-1} = \frac{\partial \ln{[R_{ch}(V_{GS})]}}{\partial V_{GS}}
\end{equation}
\fi

The nonlinear mixing of these plasma waves in the channel results in a DC-signal between source and drain electrodes that is proportional to the incident terahertz power \(P_{THz}\):
\begin{align}
\Delta V_{DS,THz} =& -\frac{V_{THz}^2}{4} \frac{\partial \ln [R_{ch}(V_{GS})]}{\partial V_{GS}}f(\omega_{THz},\tau_p)   \propto P_{THz}
\label{eq:VDSderivation}
\end{align}
\begin{align}
\Delta V_{DS,THz} &= \frac{V_{THz}^2}{4} \frac{\partial \ln [n(V_{GS})]}{\partial V_{GS}} f(\omega_{THz},\tau_p)
\label{eq:VDSderivation1}
\end{align}
where \(f(\omega,\tau_p)= 1+2\omega \tau_p/\sqrt{1+(\omega \tau_p)^2}\) is the channel's conversion ability relative to a low-frequency pure resistive mixing process \cite{zdanevicius_field-effect_2018}, and \(\tau_p\) is the momentum scattering time.\footnote{From experimental \(I\)--\(V\)-characteristics, \(\tau_p(\SI{300}{\kelvin}) \approx \qty{20}{\femto\second}\) was determined, which is in agreement with \cite{ludwig_modeling_2024}. At \qty{20}{\kelvin}, we determined \(\tau_{p}\approx\qty{46}{\femto \second}\).}
The rectified signal (\(\symrm{\Delta} V_{DS}\)) is proportional to the incident power \(P_{THz}\) within a certain linear regime.
At higher power levels, this linear response transitions into a super-linear regime (\(\symrm{\Delta} V_{DS}\propto P_{THz}^2\)), within the sub-threshold operation range before eventually going into saturation. In contrast, above the threshold, the response shifts directly from the linear regime into saturation without exhibiting the intermediate super-linear behavior~\cite{lisauskas_field-effect_2018}.

An unbiased source--drain condition (\(V_{DS,\symrm{ext}} = \SI{0}{V}\)) has been shown to yield the highest signal-to-noise ratio (SNR) compared to actively biased operation~\cite{lisauskas_terahertz_2013}.
In this unbiased case, thermal Johnson--Nyquist noise is the dominant noise source,
\begin{equation}
V_{JN} = \sqrt{4 k_B T_{Det} R_{DS}}
\label{eq:JohnsonNyquistNoise}
\end{equation}
where \(k_B\) is the Boltzmann constant, \(T_{Det}\) is detector temperature and \(R_{DS}\) is the transistor's source--drain resistance, determined experimentally from \(I\)--\(V\) characterization.
Both the enhanced rectification efficiency and the reduction of thermal noise contribute to an improved NEP under cryogenic operation.

\subsection{Detector implementation}

\begin{figure}[tb]
  \centering
  \includegraphics[keepaspectratio, width=\columnwidth]{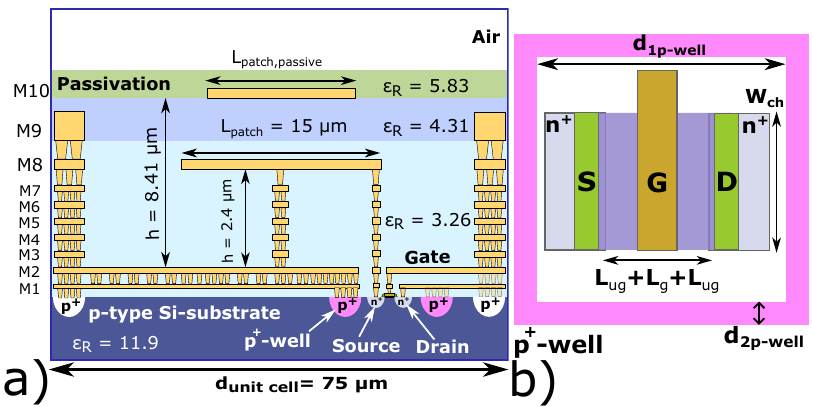}
  \caption{a)~Cross-sectional view of the patch-antenna implementation, representing the unit cell of the \(8\times8\) TeraFET array (not to scale; graphic adapted from~\cite{ludwig_modeling_2024}). Relevant dimensions and permittivities are indicated. 
  b)~Top view of the transistor layout. The strongly doped \(p^+\)-well region, shown in pink, was introduced to pin the body potential to source ground (guard ring).
  The dimensions of the square-shaped ring are \(d_{1,p\text{-}well} = \JH{\qty{6}{\micro\meter}}\) and \(d_{2,p\text{-}well} = \JH{\qty{1}{\micro\meter}}\). Depth of the \(p^+\)-well is estimated as \(\sim\)\qty{60}{\nano \meter} and doping density around \qty{2.5e19}{\centi\meter^{-3}} (compare \cite{ludwig_2-d_2025}).}  
  \label{fig:TeraFETcrossSectionB6}
\end{figure}

Two FET-based THz detector designs were investigated in this work, using TSMC's 65\mbox{-}nm foundry process (CLN65LP, type 1.2VmLowVtMOS): a single-patch antenna for \qty{540}{\GHz} operation, and an \(8\times 8\) patch-antenna array for \qty{2.85}{\THz} operation.
The 540\mbox{-}GHz detector used a patch antenna area of \qtyproduct{138x40}{\um}, as visualized in~\cite{ludwig_modeling_2024} and had a transistor channel width \(W_{ch}= \qty{1}{\um}\), and gated length \(L_{g}= \qty{60}{\nm}\).
The patch-antenna's simulated directivity is \(\sim\)\qty{5.8}{dBi}.
This leads to a low optical coupling efficiency \JH{in the experimental work described later}, as the effective antenna area is small compared to the Gaussian focal point of the \JH{off-axis parabolic (OAP) mirror with a reflected focal length (RFL) of 3-inch (Full-Width-at-Half-Maximum, \(\mathrm{FWHM} \approx\SI{1.3}{\milli \meter})\)}.
Under these focal conditions, we expect approximately \SI{5}{\percent} of the incident power to be coupled into the antenna (technique presented in \cite{ludwig_modeling_2024}).
The \qty{2.85}{\THz} detector used an \(8\times 8\) binned-element configuration with active area \qtyproduct{600x600}{\um} as reported in~\cite{holstein_88_2024}.
A cross-sectional view of the unit-cell element is visualized in \autoref{fig:TeraFETcrossSectionB6}. Transistor channel parameters \(W_{ch}= \qty{0.2}{\um}\) and gated length \qty{60}{\nm} were used. Each patch-antenna in the array had area \qtyproduct{15x15}{\um}, with the active patch placed in layer M8 (height, \(h=\qty{2.4}{\um}\) above the ground plane) as indicated in the diagram. A smaller \qtyproduct{13x13}{\um} patch is placed in the M10-layer (\(h=\qty{8.41}{\um}\)) to satisfy minimum metal fill-factor rules for foundry processing, without significantly modifying the (simulated) antenna field pattern. 
To ensure a consistent electrostatic environment and to minimize noise contributions from and to all rectifying transistors, a strongly doped \(p^+\)-well guard ring is implemented around each \(n\)-MOSFET. This ring electrically connects both the source and body terminals to the global ground potential. The high doping concentration of the \(p^+\)-well ensures sufficient conductivity even at cryogenic temperatures, thereby preserving effective grounding and suppressing latch-up effects by inhibiting the formation of parasitic current paths in the $p$-substrate.
Furthermore, the guard ring stabilizes the body potential and mitigates carrier freeze-out in the inversion layer near the source junction---an essential condition for reliable operation at low temperatures.
The dimensions of the applied square shaped ring are \(d_{1p-well} = \qty{6}{\um}\) and \(d_{2p-well} = \qty{1}{\um}\).\footnote{This technique is usually applied in the semiconductor design community to reduces noise and protect active detector element from external noise or variations in the local doping levels. The applied ring structure was already implemented in previous publications~\cite{zdanevicius_field-effect_2018,ludwig_modeling_2024}.}

\begin{figure}[tb]
  \centering
  \includegraphics[keepaspectratio, width=\FigWidth\columnwidth]{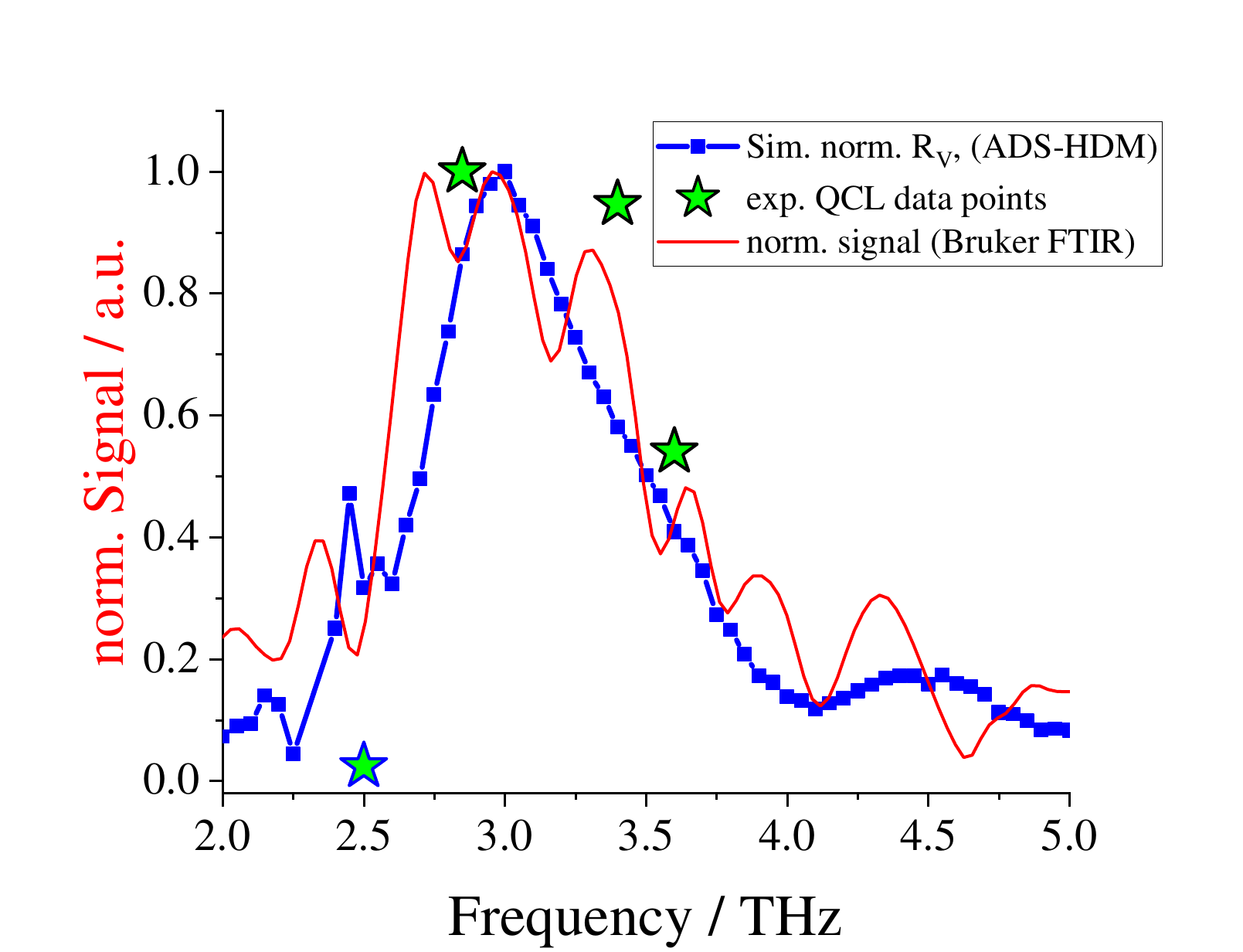}
  \caption{\JH{Experimental and simulated THz spectral response of the patch-antenna coupled FET shown in \autoref{fig:TeraFETcrossSectionB6} at \(V_{GS}=\qty{0.55}{\volt}\). Stars: Normalized responsivity taken from measurements with \(8\times8\) detector using THz-QCLs at four frequencies (\qty{2,5}{\THz},\qty{2,85}{\THz},\qty{3,4}{\THz} and \qty{3,6}{\THz}). Data at \qty{2.85}{\THz} and \qty{3.4}{\THz} was reproduced from~\cite{holstein_88_2024}. Red Line: Spectral response of a single FET-detector (unit cell of the \(8\times8\)-detector) coupled to a superstrate Si-lens as introduced in~\cite{krysl_si_2024}. Spectral  response was determined using a Michelson interferometer with a thermal (Globar) source (vendor: Bruker Corp.). Blue: Simulated responsivity of the patch antenna element. The antenna structure's impedance and efficiency was simulated using CST Studio Suite and the transistor's terahertz response was modeled using an in-house hydrodynamic model implemented in Keysight's Advanced Design System (ADS-HDM) as presented in~\cite{ludwig_circuit-based_2019} and \cite{ludwig_modeling_2024}. All measurements presented within this graph were acquired, while the  detector element was operated at room-temperature.}}  
  \label{fig:SpectralResponse}
\end{figure}

\JH{A summary of the detector element's spectral response is shown in \autoref{fig:SpectralResponse}, with continuous data obtained from a thermal source using a Bruker interferometer and discrete responsivity measurements taken using THz-QCLs at four emission frequencies (\qty{2,5}{\THz}, \qty{2,85}{\THz}, \qty{3,4}{\THz} and \qty{3,6}{\THz}). 
Results at \qty{2.85}{\THz}, \qty{3.4}{\THz} and \qty{3.6}{\THz} were acquired using THz QCLs in semi-insulating surface plasmon waveguides, with the data at \qty{2.85}{\THz} and \qty{3.4}{\THz} having been reproduced from~\cite{holstein_88_2024}.
The \qty{2.5}{\THz}-QCL data-point was acquired using a frequency rescaled design adapted from~\cite{wienold_low-voltage_2009}, and packaged into a double metal waveguide.
We attribute the relatively low signal to poor beam quality in the focal plane resulting from the double-metal waveguide.
For all measurements shown in \autoref{fig:SpectralResponse}, the respective detector element was operated at room temperature. Nonetheless, the presented data illustrate that the peak of spectral response qualitatively agrees with model predictions. The quantitative and statistical analysis requires extensive modification of THz sources, allowing formation of Gaussian beams (or a calibrated Black-Body high temperature sources, if measured using FTIR spectrometry).} 

\begin{figure}[tb]
  \centering
  \includegraphics[keepaspectratio, width=\FigWidth\columnwidth]{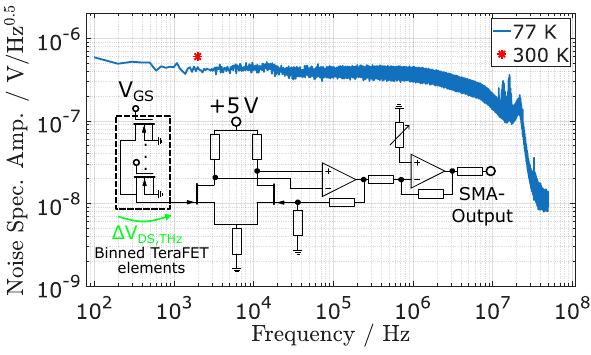}
  \caption{Experimental noise spectral amplitude of both room-temperature and \ce{LN2}-cooled \(8\times 8\) FET array operation \JH{at $V_{GS}=\qty{0.55}{\volt}$} measured with a Tektronix DPO4104 oscilloscope.
 (Inset)~Schematic of the electronic amplifier circuitry (\JH{Low Noise,  High Input Impedance Composite Amplifer}, adapted  from  \cite{texas_instruments_jfe2140_2023}) consisting of a JFET-input stage and a low-noise amplifier followed by a second low-noise amplifier stage (total gain \(G_{V}=100\)). The complete amplifier chain is not actively cooled, while the detector is operated at \qty{77}{\kelvin}. The amplifier chain is enclosed in a metal shielding box (see \autoref{fig:ln2cryo}).}
  \label{fig:ElectronicsAndNoisespectrum}
\end{figure}

\subsection{Packaging and readout electronics}
\begin{figure}[tb]
  \centering
  \includegraphics[keepaspectratio, width=\FigWidth\columnwidth]{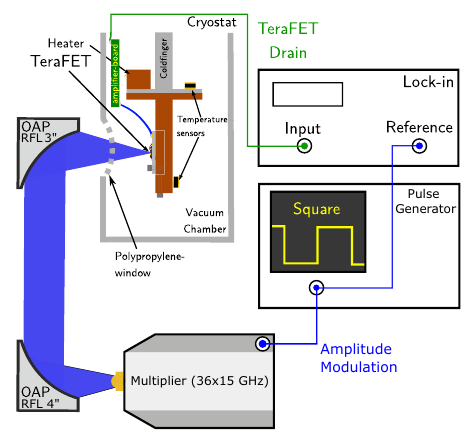}
  \caption{Experimental setup used for FET detector characterization at cryogenic temperatures at \qty{540}{\GHz}. Terahertz radiation generated using an RPG ZTX750 multiplier (\(\times\)36) (vendor: Radiometer Physics GmbH, Germany) was guided through an optical system onto the detector (without Si-lens) inside a closed cycle cryostat. For this measurement, an \JH{OAP with RFL of \qty{3}{inch}} was used as the final focusing element due to space constraints. The biasing/readout scheme was same for the 2.85~THz-QCL based setup, while the setup geometry can be seen in \autoref{fig:LeedsQCLSetup}.}  
  \label{fig:Set540ghz}
\end{figure}

The respective detectors were mounted on in-house designed \ce{Al}-based printed circuit boards (PCBs) with high thermal conductivity (\(>\)\qty{2}{\watt\per\milli\kelvin}, thickness \qty{1.5}{\mm}) using a thermally curable, two-component epoxy adhesive (EPO-TEK 353ND).
Electrical connections between the detector electrodes and the PCB were established via \ce{Al}-wire bonding.
\JH{The resulting signals were then routed via twisted pair cables from the detector PCB to separate PCBs containing the subsequent amplifier and buffer stage. The amplifier chain was operated at room temperature, while the detector-element was actively cooled by the cryostat coldfinger.
In both configurations, the electronic readout/buffer circuitry consists of \JH{junction field-effect transistor (JFET)}- and Operational Amplfier (OPA)-based components (inset in \autoref{fig:ElectronicsAndNoisespectrum}), and was designed to achieve low-noise operation and suppress \(1/f\)-noise contributions.}
First, the rectified signal is fed into a JFET-stage consisting of two JFETs, grown on the same substrate for maximized similarity in their electronic properties.
\JH{These were biased at the lowest-noise operating point (determined by the choice of the top resistors). For the temperature-dependent measurements at \SI{540}{\GHz}, the signal was buffered by a Source-follower stage as presented in~\cite{holstein_88_2024}, based on the LSK489 JFET-pair (SOT-23-6 package, vendor: Linear Systems). The buffer stage was followed by two low-noise amplifier stages (Gain 10 each, Texas Instruments OPA211/1611). 
In case of the compact \ce{LN2}-cooled cryostat, the first stage was implemented as a low-noise, high-input impedance composite amplifier, adapted from \cite{texas_instruments_jfe2140_2023}, visualized in \autoref{fig:ElectronicsAndNoisespectrum}. The input-referred noise of our low-noise, high-input impedance composite amplifier implementation was determined with \nVperHz{1,5}, which is below the thermal noise of about \SI{1}{\kilo\ohm} resistance at room temperature of \nVperHz{4,05} expected from \autoref{eq:JohnsonNyquistNoise}. 
To realize the circuit, the JFET-pair JFE2140 (SOIC-
8 package, vendor: Texas Instruments) in combination with an OPA211 followed by a second OPA211 stage was applied. In both implementations, the second amplifier stage's output was fed to a SMA connector, which was connected to the respective lab measurement equipment (compare \autoref{fig:Set540ghz}).
\JH{For \SI{540}{\GHz}-characterizations, the  total voltage gain was \(G_{V}=\num{70}\) (compare~\cite{holstein_88_2024}), while \(G_V=\num{100}\) was determined for the \ce{LN2}-configuration.}.}
All relevant electrical potentials, including the GND/Source and \(V_{GS}\), were supplied from a common in-house designed power supply PCB.
The achievable modulation bandwidth \(f_{-3dB} \approx \qty{5}{\mega \hertz}\) of the detector and amplifier electronics was estimated via measurement of noise spectral amplitude density spectrum using a Tektronix DPO4104 oscilloscope
as shown in \autoref{fig:ElectronicsAndNoisespectrum}.
Due to the low input-referred noise  of the amplifier chain, the detector's thermal noise contribution dominates to the overall noise spectral amplitude at \SI{77}{\kelvin}. We estimate the stage to add less than \qty{1}{\decibel} to the final noise spectral amplitude.

\subsection{Implemented cryogenic systems}
\label{scn:cryogenics}
\JH{Two cryogenic systems were used for this work: a closed-cycle cryostat for measurement of responsivity as a continuous function of temperature, and a \ce{LN2}-cooled cryostat for QCL power measurements.}
The detector-PCB was mechanically mounted onto the metallic coldfinger of the respective cooling system.
\JH{For the helium-based closed-cycle cryostat (vendor: LakeShore Cryotronics), precise control of the coldfinger temperature was achieved via two independent thermal sensors and a heater element. Temperature regulation was managed by a LakeShore Cryotronics Model 330 controller. Using a multiplexer circuit (TMUX136, vendor: Texas Instruments), the drain output could either be accessed directly, bypassing the amplifier electronics to determine the temperature-dependent \(I\)–\(V\) characteristic, or routed through the low-noise amplifier chain for terahertz detection experiments.}
\begin{figure}[tb]
  \centering
  \includegraphics[keepaspectratio, width=0.9\columnwidth]{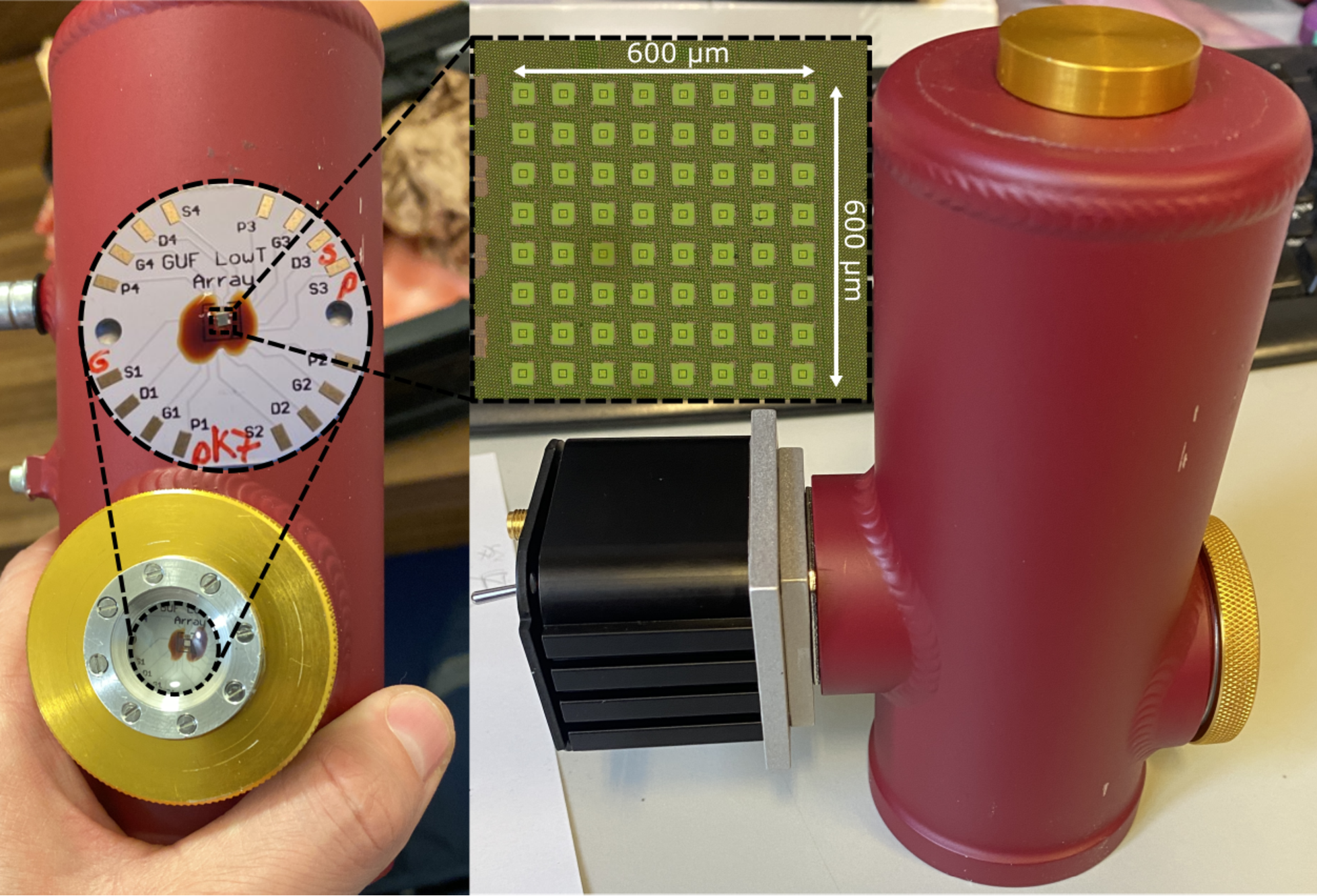}
  \caption{Photograph of the \ce{LN2}-cooled detector system and its internal components. The first inset shows the aluminum-based PCB mounted on the dewar’s cold finger, carrying the detector die. The second inset provides a zoomed-in view of the \(8\times8\) pixel binned TeraFET detector array. The resulting effective active detector area is approximately \qtyproduct{600x600}{\um} (introduced in~\cite{holstein_88_2024}). The housing provides access for liquid nitrogen cooling (top) and includes a vacuum port compatible with external pumping systems (left). 
  Buffer and amplifier electronics (see \autoref{fig:ElectronicsAndNoisespectrum}) are housed in a shielded metal enclosure to suppress electromagnetic interference from the laboratory environment. Power supply, on/off switch, and the SMA output connector are located on the rear side of the shielding box.}
  \label{fig:ln2cryo}
\end{figure}
The compact \ce{LN2}-cooled system, designed for QCL power measurements, was based on a commercial dewar system (vendor: PerkinElmer) providing both easy-access cooling and vacuum capabilities and is shown in \autoref{fig:ln2cryo}.
To enable optical terahertz transmissive access, the original \ce{CaF2} window was replaced by a \SI{20}{\micro\meter}-thick THz-transmissive polypropylene foil stretched within an in-house designed holder to ensure proper vacuum sealing.
\JH{The amplifier electronics were housed in a metallic enclosure at the rear of the dewar.}
\JH{We determined an approximate cooling time of \qtyrange{1.5}{2}{\hour} for the detector system to reach thermal equilibrium with the \qty{77}{\kelvin} coldfinger, as measured by the stabilization time of \(I_{DS}\) under a constant bias of \(V_{DS} = \qty{1}{\milli\volt}\) and \(V_{GS} = \qty{0.55}{\volt}\).}

\newpage
\section{Low-temperature responsivity analysis}
\JH{A comparable experimental apparatus was used for FET-detector characterization at \qty{540}{\giga \hertz} and \qty{2.85}{\THz}, albeit with different THz sources in each case. \qty{540}{\giga \hertz}-radiation was generated by multiplying a \qty{15}{\GHz} oscillator signal by (\(\times 36\)) using an RPG ZTX750 multiplier (vendor: Radiometer Physics GmbH, Germany). The emitted signal was amplitude modulated using an input square-wave reference for lock-in detection. The available \(P_{THz}=\qty{50}{\micro \watt}\) was measured using a pyroelectric detector (THz 10 HS, vendor: SLT GmbH~\cite{slt_sensor-_und_lasertechnik_gmbh_ptb-calibrated_2025}), calibrated by Physikalisch-Technische-Bundesanstalt (PTB,Braunschweig, Germany).}

\begin{figure}[ht]
  \centering
  \includegraphics[keepaspectratio, width=\FigWidth\columnwidth]{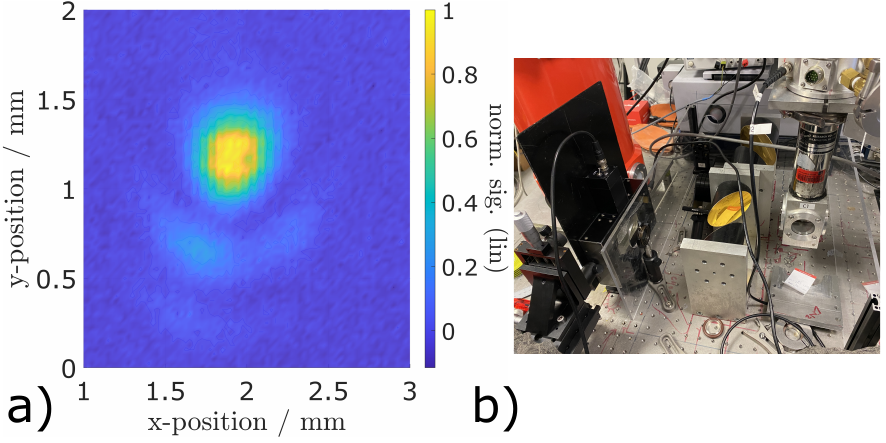}
  \caption{\JH{(a)~Normalized experimental beam profile data taken at the focal position, determined using an uncooled microbolometer camera (Swiss Terahertz Rigi S2, \qtyproduct{160x120}{px} at \qty{25}{\micro\meter} pitch). (b)~Picture of the experimental setup used for detector characterization. Emission from a THz-QCL mounted in a cryostat is guided through off-axis parabolic mirror optics (Diameter: \qty{3}{inch}, RFL: \qty{7}{inch}). The experimental beam focus is marked with an iris. The terahertz power \(P_{\symrm{THz}}\) was measured at the focal position using a Thomas Keating photoacoustic power meter.}}  
  \label{fig:LeedsQCLSetup}
\end{figure}

\JH{The \qty{2,85}{\THz} measurements were performed using a QCL adapted from the design in~\cite{wienold_low-voltage_2009}, as in our previous work~\cite{holstein_88_2024}. The QCL was mounted in a helium flow cryostat (vendor: Janis/LakeShore Cryotronics) equipped with 3\mbox{-}mm-thick HDPE windows. The QCL temperature was set at \qty{20}{K}, while being driven with either 700~mA or 808~mA pulses, with 5~\% and 50~\% duty cycle pulses generated by laboratory pulse generator (Agilent 8114A). The corresponding THz power was determined using a calibrated photoacoustic power meter (vendor: TK Instruments Ltd., UK) at the detector position shown in \autoref{fig:LeedsQCLSetup}.
}
\JH{In  case of \SI{50}{\percent} duty-cycle pulses, a QCL output power of \(P_{THz}(\qty{808}{\milli \ampere}) = \SI{2.1}{\milli\watt}\) was measured, whereas for \qty{5}{\percent}-duty-cycle operation, \(P_{THz}(\qty{700}{\milli \ampere})=\qty{0,36}{\milli\watt}\) and \(P_{THz}(\qty{808}{\milli \ampere})=\qty{0.43}{\milli\watt}\) were measured.\footnote{The corresponding peak-powers during the \qty{5}{\percent} on-time are \(\widehat{P}_{THz}(\qty{700}{\milli \ampere})=\qty{3,6}{\milli\watt}\) and \(\widehat{P}_{THz}(\qty{808}{\milli \ampere})=\qty{4,3}{\milli\watt}\) as the power meter is referenced to a \qty{50}{\percent} duty cyle.}}
\JH{The QCL's emission spectrum was found to be a single mode at \qty{2.85}{\THz}, using a commercial FTIR-system (Bruker IFS/66).
For the \qty{540}{\giga \hertz} setup, the detector was initially characterized as a function of temperature using the closed-cycle cryocooler system apparatus described in \autoref{scn:cryogenics} and shown in \autoref{fig:Set540ghz}. The radiation emitted by the respective THz-source was guided through a quasi-optical system consisting of two \qty{90}{\degree} OAPs onto the respective detector. For the \qty{540}{\GHz}-system, RFLs of \qty{4}{inch}/\qty{3}{inch} and \qty{2}{inch}-diameter were used (see \autoref{fig:Set540ghz}), while the \qty{2,85}{\THz}-system used OAPs with a diameter of \qty{3}{inch} and RFLs of \qty{7}{inch}/\qty{7}{inch} (see \autoref{fig:LeedsQCLSetup}).
For the \SI{540}{\GHz} system, the single patch-antenna coupled detector was mounted within a continuously temperature tunable closed-cycle cryostat.
For the \SI{2.85}{\THz}-characterization, the \(8\times8\) detector was operated inside the liquid nitrogen cooled cryostat.}

\JH{After passing the respective amplifier stage with voltage gain \(G_V\), the terahertz rectified voltage was demodulated using an Ametek DSP 7265 lock-in amplifier and its time-averaged amplitude \(V_{DS,LIA}\) was recorded. Voltage-responsivity \(\symcal{R}_{V,opt}\) was calculated via 
\begin{equation}
  \symcal{R}_{V,opt} = \frac{\alpha_{LIA}\cdot \symrm{\Delta} V_{DS,LIA}}{P_{THz}\cdot G_V}
  \label{eq:voltageResponsivityopt}
\end{equation}
where \(\alpha_{\symrm{LIA}} = \pi/\sqrt{2}\) accounts for the square-wave modulated signal sensed via lock-in detection and \(P_{\symrm{THz}}\) denotes the total available THz power incident on the detector.}

\JH{An alternative metric, the cross-sectional responsivity \(\symcal{R}_{V,CS}\), relates the effective antenna area to the physical dimensions of the focal spot as described in \cite{ikamas_all-electronic_2021}. 
For the single-patch antenna, the effective antenna area is 
\(A_{eff} = D\lambda_{THz}^2/(4\pi)\), where \(D\) is the simulated antenna directivity.\footnote{\(D=\qty{5.8}{dBi}\) for the \qty{540}{\GHz} patch-antenna, in good agreement with \cite{krysl_si_2024}.}}
\JH{The experimental beam profile in the focal plane of the \SI{2,85}{\THz}-setup is shown in \autoref{fig:LeedsQCLSetup}. It was acquired using an uncooled microbolometer camera (Swiss Terahertz Rigi S2, \qtyproduct{160x120}{px} at \qty{25}{\micro\meter} pitch). The focal dimensions of the beam profile were measured as \(\symrm{FWHM}_x=\qty{0.40}{\milli \meter}\), \(\symrm{FWHM}_y=\qty{0.41}{\milli \meter}\) using a numerical fitting routine. The presented beam profile was measured at \(I_{QCL}=\qty{700}{\milli \ampere}\) using 50~\% duty-cycle pulses.\footnote{Higher \(I_{QCL}\) caused saturation of the camera-pixels.}}

\JH{The optical NEP was calculated from \autoref{eq:voltageResponsivityopt} and \autoref{eq:JohnsonNyquistNoise} via
\begin{equation}
\symrm{NEP}_{opt/CS,JN/exp} = \frac{V_{JN}}{R_{V,opt/CS}} \approx \frac{V_{N,exp}}{R_{V,opt/CS}}
\label{eq:CompNEPopt}
\end{equation}
In \autoref{eq:CompNEPopt}, \(V_{N,\symrm{exp}}\) denotes the experimentally measured noise spectral density under the given laboratory conditions. 
Although we observe \(V_{JN} \approx V_{N,\symrm{exp}}\) at room temperature, as suggested in~\cite{lisauskas_terahertz_2013}, deviations from this expectation become noticeable at \ce{LN2} temperatures. We attribute these deviations to the fact that, in our experimental setup, the readout electronics remain at room temperature while the detector temperature \(T_{\symrm{Det}}\) is cooled to cryogenic temperatures (e.g., \qty{77}{K}). As it is common practice in the literature to report NEP\(_{\symrm{exp},\symrm{JN}}\), both values will be provided in the following discussion.} 
\JH{Another important Figure of Merit is experimental signal-to-noise ratio (SNR) for a given lock-in integration time \(T_C \approx 1/(4 \symrm{\Delta} f)\), where \(\symrm{\Delta} f\) is the equivalent noise bandwidth (ENBW),\footnote{The previously stated formula holds true for \qty{6}{dB\per octave} filter slope, which was applied during the experiments. For different filter slopes, \(\symrm{\Delta} f\) can be read from~\cite{ametek_advanced_measurement_technology_inc_model_2002}.} calculated via
\begin{equation}
\symrm{SNR} = \frac{\symrm{\Delta} V_{DS,LIA,THzon}}{\symrm{\Delta} V_{DS,LIA,THzoff}} = \frac{P_{THz}}{\symrm{NEP}_{opt} \cdot \sqrt{\symrm{\Delta} f}}
\label{eq:SNR_NEPDefinition}
\end{equation}
The respective dynamic range can be defined via: 
\begin{equation}
\symrm{DNR(T_{Det})} = \underset{V_{GS},P_{THz},\Delta f}{\max} \left( \symrm{SNR} \right)
\label{eq:DNR_Definition}
\end{equation}}

\section{Results}

\subsection{Temperature dependent study at 540~GHz}
\begin{figure}[tb]
  \centering
  \includegraphics[keepaspectratio, width=\FigWidth\columnwidth]{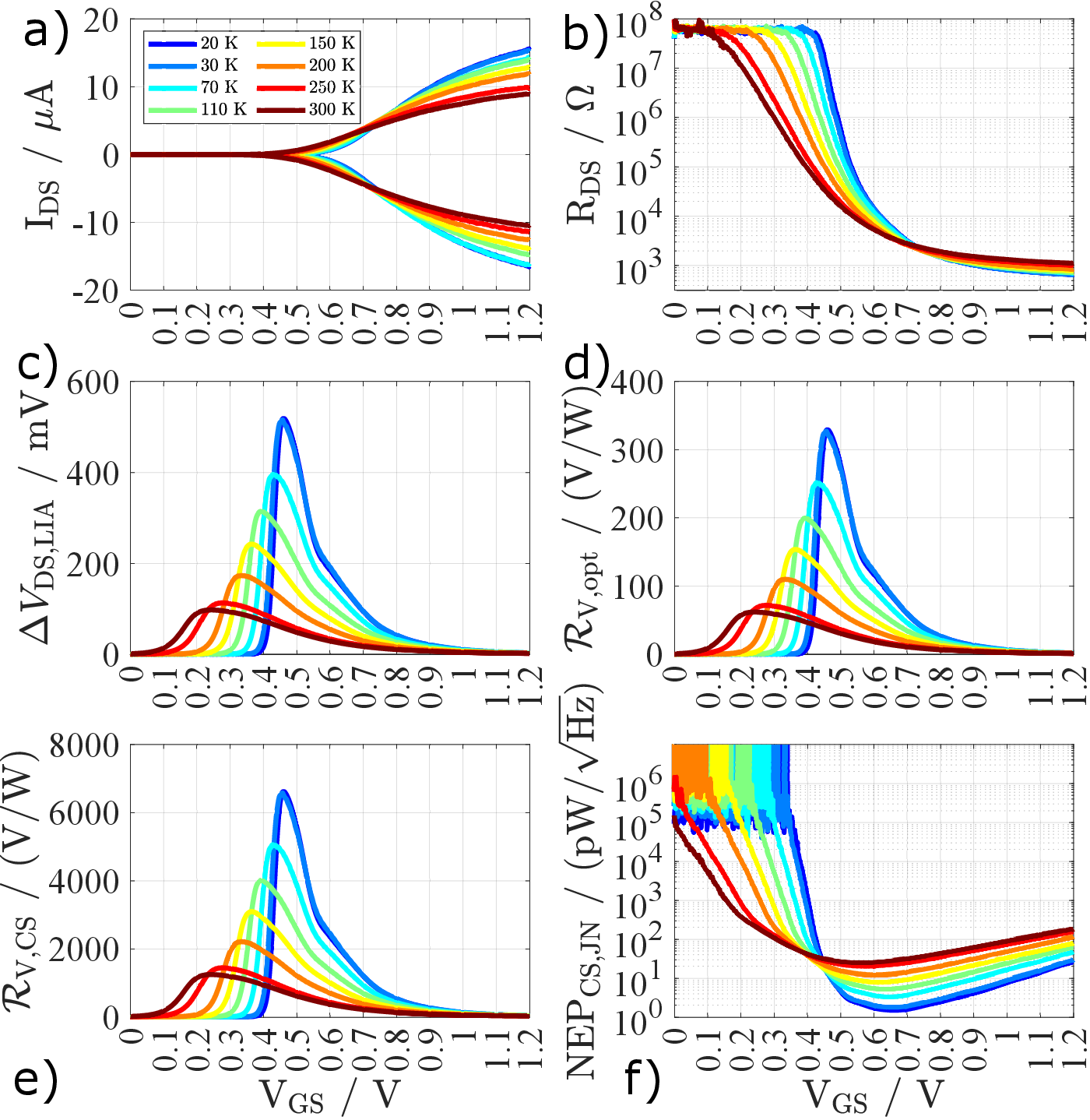}
  \caption{(a)~\(I\)--\(V\) characteristics of single-patch detector in a  \qtyrange{20}{300}{\kelvin} temperature range, measured for \(V_{DS}\pm\SI{10}{\milli \volt}\). (b)~The \(R_{DS}(V_{GS},T_{Det})\)-behavior, calculated from the \(I\)--\(V\) characteristics. 
  (c--d)~Rectified raw terahertz signal measured with a LIA at focal center position and corresponding optical responsivity \JH{at \SI{540}{\GHz}}.
  (e--f)~Cross-sectional responsivity and corresponding \JH{\(\symrm{NEP}_{CS,JN}\).}}  
  \label{fig:Res540ghz}
\end{figure}

\autoref{fig:Res540ghz} summarizes the results of a \(T_{Det}\)-dependent study of patch-antenna coupled detectors resonating at \qty{540}{\giga \hertz}. 
\JH{Down to \qty{20}{\kelvin}, we do not see indications of strong device degradation due to carrier freezeout-related effects - which is often observed in Si-MOSFETs in this temperature range - in both IV-characteristic and THz-responsivity. We attribute this effect primarily to efficient substrate potential-pinning by the strongly doped $p^+$-well placed in close proximity around the $n$-MOSFET (see \autoref{fig:TeraFETcrossSectionB6}).}
In addition, the \mbox{\(I\)--\(V\)}-characteristic shows the Zero-Temperature-Coefficient (ZTC) point (see~\cite{saligram_future_2024}) at approximately \SI{0.7}{\volt}.
Above this bias point, a temperature-independent \(R_{DS}\) can be observed.
NEP continuously improves (increases) as the temperature reduces, as a result of decreasing (thermal) noise contributions and increasing responsivity \(\symcal{R}_{V,CS}\).
At \qty{20}{\kelvin}, a relative decrease in \(\symrm{NEP}_{CS,JN}\) \JH{by a factor of 11 to 15} compared to room-temperature operation was observed in the characterization of three independent detectors.
At \qty{77}{\kelvin}, we found a relative decrease in \(\symrm{NEP}_{CS,JN}\) by a factor of 4--6, compared to room-temperature, which is consistent with previously reported results for Si-CMOS TeraFETs implemented in 90-nm technology~\cite{ikamas_optical_2020}.\footnote{First indications of carrier freeze-out effects were found around \qty{9}{\kelvin} (dataset not shown here).
At this temperature, quantization effects of the two-dimensional electron gas (2DEG) become visible in the IV-characteristics within the sub-threshold region.}
\JH{One should note that the approximately fourfold improvement in NEP around the boiling temperature of \ce{LN2} can be predicted by applying \autoref{eq:VDSderivation} to the DC data presented in \autoref{fig:Res540ghz}.\footnote{The temperature dependent scattering time  \(\tau_p(T)\), relevant for \(f(\omega_{THz},\tau_p)\), was determined by fitting \(R_{DS}(V_{GS},T)\) within the UCCM \cite{shur_unified_1992}.} For the \qty{20}{\kelvin} results, \autoref{eq:VDSderivation} predicts an NEP reduction factor of \(\sim\)\num{8}, while the experimental reduction factor was \(\sim\)\num{14}.}

\subsection{Nitrogen cooled system at 2.85~THz}
\begin{figure}[tb]
  \centering
  \includegraphics[keepaspectratio, width=0.9\columnwidth]{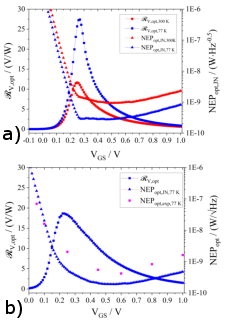}
  \caption{\JH{Optical} voltage responsivity and NEP of detector system as a function of \(V_{GS}\), using a \qty{2.85}{\tera \hertz}-QCL source operating at \(T_{QCL}=\qty{20}{\kelvin}\).
  (a)~Room-temperature  vs. \ce{LN2}-cooled detector operation, with QCL parameters: \qty{5}{\percent}-duty cycle, \(f_{mod}=\qty{10}{\kilo \hertz}\), \(I_{QCL,on}=\qty{700}{\milli \ampere}\)). (b)~Cooled detector operation with QCL parameters: \qty{50}{\percent} duty cycle, \(f_{mod}=\qty{2}{\kHz}\), \(I_{QCL,on}=\qty{808}{\milli \ampere}\)). The experimental noise at different \(V_{GS}\)-potentials was determined under the same laboratory conditions.}  
  \label{fig:PerformanceGateSweeps}
\end{figure}

\begin{figure}[tb]
  \centering
  \includegraphics[keepaspectratio, width=\FigWidth\columnwidth]{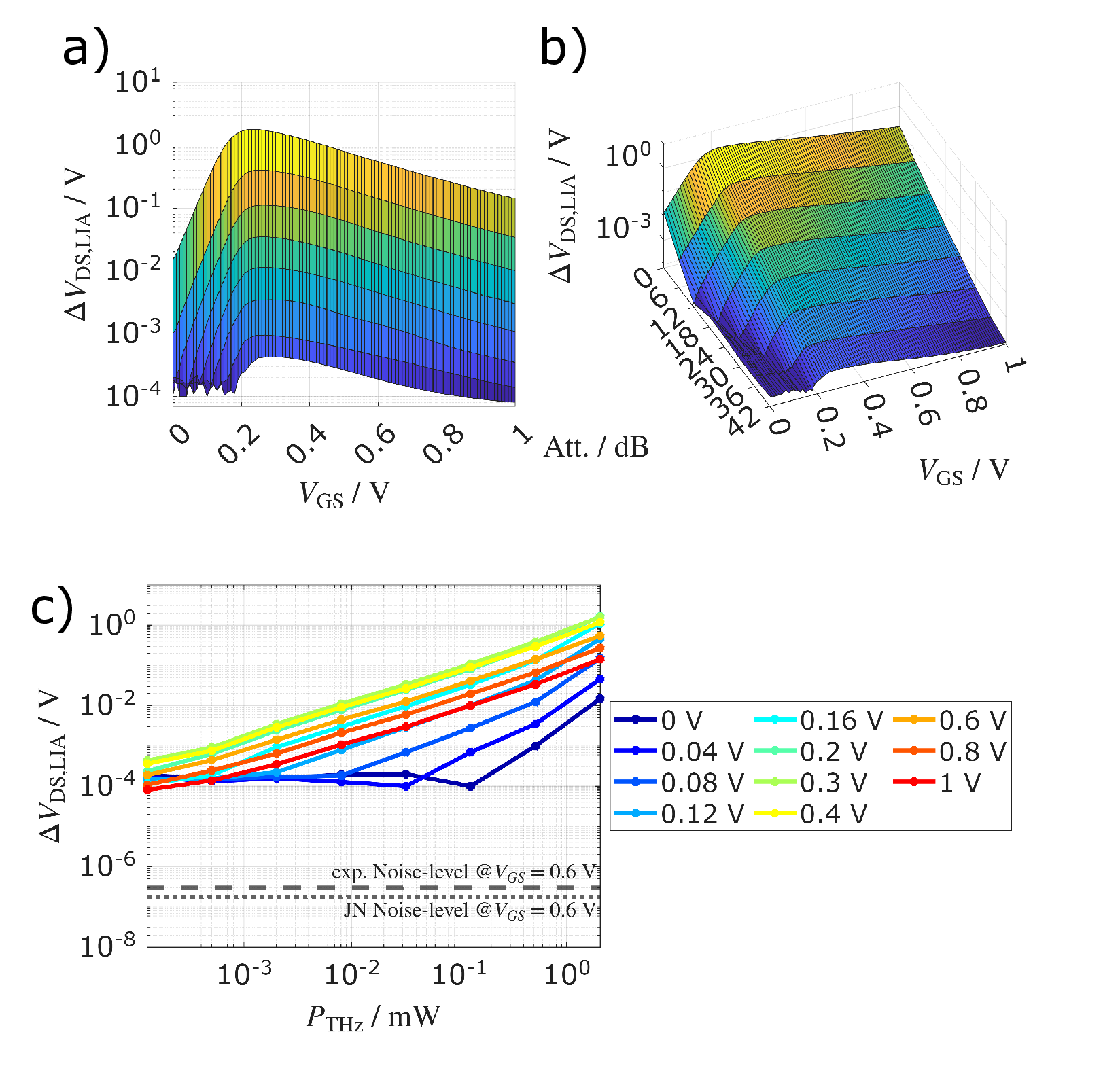}
  \caption{(a--b)~Investigation of detector linearity behavior. \(V_{GS}\)-dependent \(\Delta V_{DS,LIA}\) (raw data, not corrected for amplifier gain \(G_V\)) as function of power attenuation (QCL-operation parameters: \qty{50}{\percent}-duty cycle, \(f_{mod}=\qty{2}{\kHz}\), \(I_{QCL,on}=\qty{808}{\milli \ampere}\),  \(T_{QCL}=\qty{20}{\kelvin}\)). (c)~Rectified \(\Delta V_{DS,LIA}\) for differing  \(V_{GS}\) indicating linear operation \(\Delta V_{DS,LIA} \propto P_{THz}^1\) for \(V_{GS}>\SI{0.2}{\volt}\). For \(V_{GS}<\qty{0.2}{\volt}\), \JH{super-linear}  \(\Delta V_{DS,LIA} \propto P_{THz}^2\)-behavior occurs.}  
  \label{fig:LinearityCheck}
\end{figure}

\autoref{fig:PerformanceGateSweeps} shows the detector's optical responsivity and NEP as function of \(V_{GS}\) at the focus of a \qty{2.85}{\THz}-QCL (\qty{5}{\percent} duty-cycle pulses).
The NEP is shown to decrease by a factor of 3.5 (\nep{950} vs. \nep{270}) when the detector is cooled from \qty{300}{K} and \qty{77}{K}. 
\JH{\autoref{fig:PerformanceGateSweeps}(b) presents the results obtained using a \qty{50}{\percent} duty-cycle pulse. In addition to \(\symrm{NEP}_{JN,opt}\), we report the experimental optical NEP (\(\symrm{NEP}_{exp}\)), calculated from the total system noise measured under laboratory conditions.
The minimum values are \(\symrm{NEP}_{opt,JN} \approx \nep{190}\) and \(\symrm{NEP}_{opt,exp} \approx \nep{420}\).
The minimum \(\symrm{NEP}_{opt,exp}\) corresponds to an experimental DNR of approximately \qty{64}{dB} (\(P_{\symrm{THz}}=\qty{2.1}{\milli\watt}\), \(T_{\mathrm{C}}=\qty{100}{\milli\second}\)) and up to \qty{67}{dB} for a detection bandwidth of \(\Delta f=\qty{1}{\hertz}\).
As shown in \autoref{fig:LinearityCheck}, a DNR of approximately \qty{70}{dB} is expected under idealized noise conditions for \(\Delta f=\qty{1}{\hertz}\).
When directly comparing the results presented in \autoref{fig:Res540ghz} and \autoref{fig:PerformanceGateSweeps}, an important characteristic of the \(8\times8\) pixel-binned detector must be considered (see the derivation in~\cite{holstein_88_2024}).
For \(N = 64\) pixels combined in parallel, the voltage responsivity is reduced by a factor of 64 relative to the cross-sectional responsivity of a single pixel. In contrast, the NEP is expected to increase only by a factor of \(\sqrt{N}=8\), due to the reduced thermal noise contribution resulting from the parallel network configuration.}

A study of detector linearity as a function of \(V_{GS}\) for \qty{50}{\percent}-duty-cycle QCL pulses is presented in \autoref{fig:LinearityCheck}.
The THz power (\(P_{THz}=\qty{2.1}{\milli \watt}\)) impinging on the detector was attenuated in \qty{6}{dB} steps to a maximum of \qty{42}{dB} attenuation (\(P_{THz}\approx\qty{125}{\nano \watt}\)).
Due to the high experimental dynamic range exceeding the available dynamic range of the lock-in amplifier (\(\sim\)\qty{40}{dB}), \(\Delta V_{DS,\symrm{LIA},\symrm{THz,off}}\) was determined using a lower sensitivity setting, which allowed resolution of the corresponding noise level. 
\JH{The} detector shows a linear response for \(V_{GS} > \qty{0.2}{\volt}\), while a super-linear response (\(\Delta V_{DS} \propto P_{THz}^2\)) occurs for \(V_{GS} <\qty{0.2}{\volt}\). This super-linear response can occur in \JH{FET-based} detectors within the sub-threshold regime as a precursor to saturation, whereas in the above-threshold regime, the detector typically transitions directly into saturation. Due to this nonlinear behavior, a FET operated in this regime can, for example, be used as an autocorrelator~\cite{ikamas_sub-picosecond_2018,lisauskas_field-effect_2018,ikamas_silicon_2018}.
\JH{Extrapolation techniques can be employed to determine the minimum detectable power level and the corresponding SNR, which are in good agreement with the prediction based on the experimentally derived NEP (see \autoref{eq:SNR_NEPDefinition}). The results further indicate a linear detector response at the recommended operating bias point, approximately \qty{0.6}{\volt}, corresponding to the minimum NEP.}

\section{Summary and Outline}
\tiny
\begin{table*}[htbp]
  \centering
  \caption{Overview of terahertz power \JH{detector systems} around \qty{3}{\tera \hertz}.}
  \rowcolors{1}{gray!25}{white} 
    \centering
   \begin{tabularx}{\textwidth}{|m{6em}|m{9.5em}|m{8em}|m{6em}|m{5em}|m{6em}|m{6em}|m{6em}|}
   \toprule
\textbf{Property/ THz-detector} & \textbf{FET \(8\times 8\) (65\mbox{-}nm  Si-CMOS)}& \textbf{FET (AlGaAs/GaAs)} & \textbf{Schottky Diode} & \textbf{NbTES bolometer} & \textbf{InSb HEB} & \textbf{Pyroelectric detector} & \textbf{Golay cell} \\
\textbf{Frequency range} & here \qtyrange{2.8}{3.6}{\THz} (Intrinsically \qtyrange{0.1}{30}{\THz}, adjustable via different antenna designs) & \qtyrange{0.1}{30}{\THz} covered by broadband antenna (log-spiral) & power detection below \qty{1.5}{\THz}, above applications as mixers  & \qtyrange{0.1}{30}{\THz}, lowpass filters exist & \qty{60}{\GHz} to \qty{2.5}{\THz} (\(>\)\qty{0.5}{\THz} only with magnetic field) & up to IR range & GHz up to IR range  \\
\(f_{mod,-\qty{3}{\dB}}\) & DC--\qty{5}{\MHz} (limited by readout electronics) & not stated (see  comment). & up to \qty{40}{\GHz} & \qty{1}{\kHz} & \qty{0.5}{\MHz} & \qty{200}{\Hz} & \qty{20}{\Hz} \\
\textbf{\JH{DC operation}} & yes  & yes & yes, strong \(1/f\) noise  & no & no information & no  & no \\
\(\symbf{\symrm{NEP}_{opt}}\) & \makecell{\nep{190} at \qty{77}{\kelvin} \\ exp. \nep{420}}  & \nepnW{453} at \qty{3.08}{\THz}  & \nep{10} below \qty{1}{\THz}. Roll-off above & \nep{1} & \nep{1.5}& \nepnW{1} to \nepuW{1} & \nep{200} \\
\textbf{Linear operation range} & \qty{0.4}{\nano \watt} to \qty{2}{\milli \watt} \(>\)\qty{67}{dB}, superlinear response available via \(V_{GS}\)-variation (e.g. for autocorrelation methods) & up to \qty[separate-uncertainty=true]{69\pm3}{\dB}/\qty{1}{\dB} compression point at \qty{0,6}{\kilo\watt} at \qty{1,3}{\THz}.  & around \qty{60}{\dB} & \qty{1}{\pico \watt} to \qty{10}{\micro \watt} (\qty{70}{\decibel}) & no information & \qty{1}{\micro \watt} to \qty{5}{\milli \watt} (\qty{37}{\decibel}) & \qty{0.2}{\nano  \watt} to \qty{100}{\micro \watt} (\qty{60}{\decibel}) \\
\textbf{Operation Temperature} &  Operable \qtyrange{9}{300}{K} (RT), here \qty{77}{K} (we expect operability towards lower temperatures, \qty{9}{K} is limitation of cooling system) & RT, \qty{20}{\kelvin} operation shown in~\cite{klimenko_temperature_2012}  & RT & \qty{4}{K}  & \qty{4}{K} & RT & RT \\
\textbf{Coupling optics/polarisation} & None, Linear polarisation. Optical coupling depends on optical mode profile. Optical elements can be applied to increase performance & Si-lens/Horn-antenna  & Waveguide integration/Si lens coupled to antenna. Polarisation dependent. & Winston cone/no polarisation & Winston cone/no polarisation & None/no polarisation & None/no polarisation \\
\textbf{Detection mechanism} & Plasma wave & Plasma wave & Rectification in Diode & electro-thermal feedback at SC-transition (\qty{8}{K}) & Hot-electron phenomenon & thermal & thermal \\
\textbf{Comment} & Detection speed currently limited by applied readout electronics. Spectral response mainly depends on antenna design. Intrinsic effect on \qtyrange{10}{20}{\pico\second} timescale. & Implementations mostly applied for high power Free-Electron-Laser (FEL) sensing applications mapping trace of the pulse down to \(\tau_{FWHM}=\qty{24}{\pico \second}\) (limited by \qty{30}{\giga \hertz} oscilloscope). & Mostly applied as heterodyne mixers above \qty{1.5}{\THz}. NEP not stated above \qty{1.7}{\THz}  & commercial & commercial & commercial, absolute power  calibration available \cite{slt_sensor-_und_lasertechnik_gmbh_ptb-calibrated_2025} &  commercial \\
\textbf{References} & \cite{holstein_88_2024}, \textbf{This work} &  \cite{yadav_gaas-based_2024,regensburger_broadband_2015}  & \cite{mehdi_thz_2017,virginia_diodes_inc_zero_2025} & \cite{qmc_instruments_ltd_superconducting_2025} & \cite{qmc_instruments_ltd_insb_2025}  & \cite{slt_sensor-_und_lasertechnik_gmbh_ptb-calibrated_2025,bielecki_review_2024} & \cite{tydex_golay_2025,bielecki_review_2024} \\
\bottomrule
\end{tabularx}%
  \label{tab:addlabel}%
\end{table*}%
\normalsize
\JH{We have demonstrated a compact, liquid-nitrogen-cooled FET-based THz detector system with strong potential for sensing applications in the \qtyrange{2.8}{3.6}{\THz} range.
Compared with helium-cooled systems, liquid-nitrogen cooling introduces less operational overhead and is more readily suited to both laboratory and field deployments.
Cooling at \qty{77}{\K} is also more readily achievable in satellite instrumentation payloads than \qty{4}{\K} stages.
Furthermore, the underlying detector technology can be adapted to operation across a wide range of THz frequencies through application-specific antenna designs.}
Our results confirm reliable detector operation across a broad temperature range, from room temperature down to at least \qty{20}{\K}, without observable carrier freeze-out effects --- phenomena that are typically expected to limit or inhibit the operation of semiconductor-based detectors at cryogenic temperatures. This thermal robustness presents a key advantage over many conventional THz detectors, which are mostly restricted to narrow temperature windows. Moreover, the detector exhibits continuous improvement in sensitivity (in agreement with the literature) with decreasing temperature, allowing the system to be tailored to the dynamic range requirements of specific experimental conditions.

Although the noise-equivalent power (NEP) of the FET-based system is higher than that of state-of-the-art bolometric detectors, it achieves a comparable linear dynamic range of approximately \qty{67}{\dB} (\(\Delta f = \qty{1}{\hertz}\)) when used with high-power THz QCL sources (\(P_{THz} = \qty{2.1}{\milli\watt}\)).
The full dynamic range at the operating point (\(V_{GS}=\qty{0.6}{\volt}\)) could be achieved with the available power, as the detector was still operating linearly. In contrast, highly sensitive bolometers such as Nb-TES typically require attenuation when exposed to similar power levels (e.g., linear regimes between \qty{1}{\pico \watt} and \qtyrange{10}{20}{\micro\watt} corresponding to approx \qty{70}{\dB} linear dynamic range).
In this context, we emphasize that the reported detector sensitivity is reduced by approximately a factor of 8 due to pixel binning -compared to the respective single detector element under assumption of full power coupling- which was applied to increase the active detector area and achieve higher readout speeds. This approach enables practical, real-world sensing applications without the need for additional coupling elements such as silicon lenses or horn antennas. For resonant single-element TeraFET detectors-typically used in combination with silicon lenses-optical NEPs below \nep{20} have been demonstrated at room temperature in the sub-\SI{1}{\THz} range. 
Based on the results previously presented for the sub-1-THz range, we expect the optical NEP of deep-cryogenically (down to 20~K) cooled TeraFETs to reach values \JH{of approx. \nep{1.8}}, comparable to state-of-the-art bolometers, while retaining the FET-detector’s inherently fast detection speed, broad operational temperature range and spectral specificity --- properties that are crucial for future terahertz (gas) spectroscopy applications. However, in the frequency range around \SI{3}{\tera\hertz}, the NEP behavior at cryogenic temperatures approaching \SI{20}{\kelvin} remains to be investigated in future studies. In the present work at \SI{77}{\kelvin}, optimized for practical applications we observed a relative NEP improvement by a factor of 3.5 compared to room-temperature operation, which is in good agreement with the relative improvement reported in the sub-1~THz range.
Unlike conventional (mostly thermal) detectors operating in the 3~THz range (see \autoref{tab:addlabel}), which are typically limited to modulation frequencies in the kilohertz range, the realized TeraFET-based system supports significantly higher modulation speeds. In its current implementation, the system bandwidth is primarily limited by the low-noise readout electronics, which are optimized for wideband operation up to several megahertz and for effective suppression of $1/f$ noise (see \autoref{fig:ElectronicsAndNoisespectrum}).
\JH{On the one hand, this makes the system well suited for DC-true terahertz detection, such as continuous-wave (cw) power monitoring of unmodulated terahertz sources (e.g., THz-QCLs~\cite{holstein_comparison_2025} used as local oscillators (LOs) in heterodyne receivers). A DC-true detection scheme refers to reading out the DC rectified voltage (e.g., using a multimeter), which is proportional to the incident terahertz power. This approach reduces the overall instrumentation complexity because no lock-in amplifier is required (compare \cite{zdanevicius_field-effect_2018}).}  
On the other hand, the high bandwidth enables time-resolved terahertz gas spectroscopy, capable of tracking transient phenomena on sub-\SI{}{\micro\second} timescales. In addition to its demonstrated high linear dynamic range ($>$64~dB,$T_C=\SI{100}{\milli \second}$), the detector can be switched between linear and super-linear (quadratic) response regimes simply by adjusting the gate voltage. This operational flexibility offers significant versatility, making the system suitable for a wide range of experimental applications, including autocorrelation measurements.

\section*{Acknowledgments}
\JH{We thank Dr. Florian Ludwig from Institut Català de Nanociència i Nanotecnologia (ICN2), Barcelona,  Spain, previously Physikalisches Insitut (U Frankfurt)} for providing access to the ADS-HDM model. We also thank the staff of the institute’s mechanical workshop for their constructive collaboration and for manufacturing custom components used in the cryogenic detector characterization setups.

\section*{Data Availability}
Data associated with this paper are freely available at the Zenodo Data Repository DOI: \url{ https://doi.org/10.5281/zenodo.15865221}

\section*{Author contributions}
J.~Holstein: Investigation (lead), Methodology, Software, Validation, Visualization, Data Curation, Writing --- original draft;
N.~K.~North: Investigation, Writing --- Review and Editing;
A.~Hof: Investigation;
S.~S.~Kondawar: Investigation;
D.~B.~But: Investigation;
M.~Salih: Investigation;
L.~Li: Investigation, Resources;
E.~H.~Linfield: Conceptualization, Funding Acquisition;
A.~G.~Davies: Conceptualization, Funding Acquisition;
J.~R.~Freeman: Conceptualization, Funding Acquisition;
A.~Valavanis: Conceptualization, Methodology, Investigation, Funding Acquisition, Supervision, Writing --- Review and Editing;
A.~Lisauskas: Conceptualization, Methodology, Funding Acquisition, Supervision;
H.~G.~Roskos: Conceptualization, Methodology, Funding Acquisition, Supervision, Writing --- Review and Editing.

\ifCLASSOPTIONcaptionsoff
  \newpage
\fi

\bibliography{JHbib}
\bibliographystyle{IEEEtranDOI}

\begin{IEEEbiography}[{\includegraphics[width=1in,height=1.25in,clip,keepaspectratio]{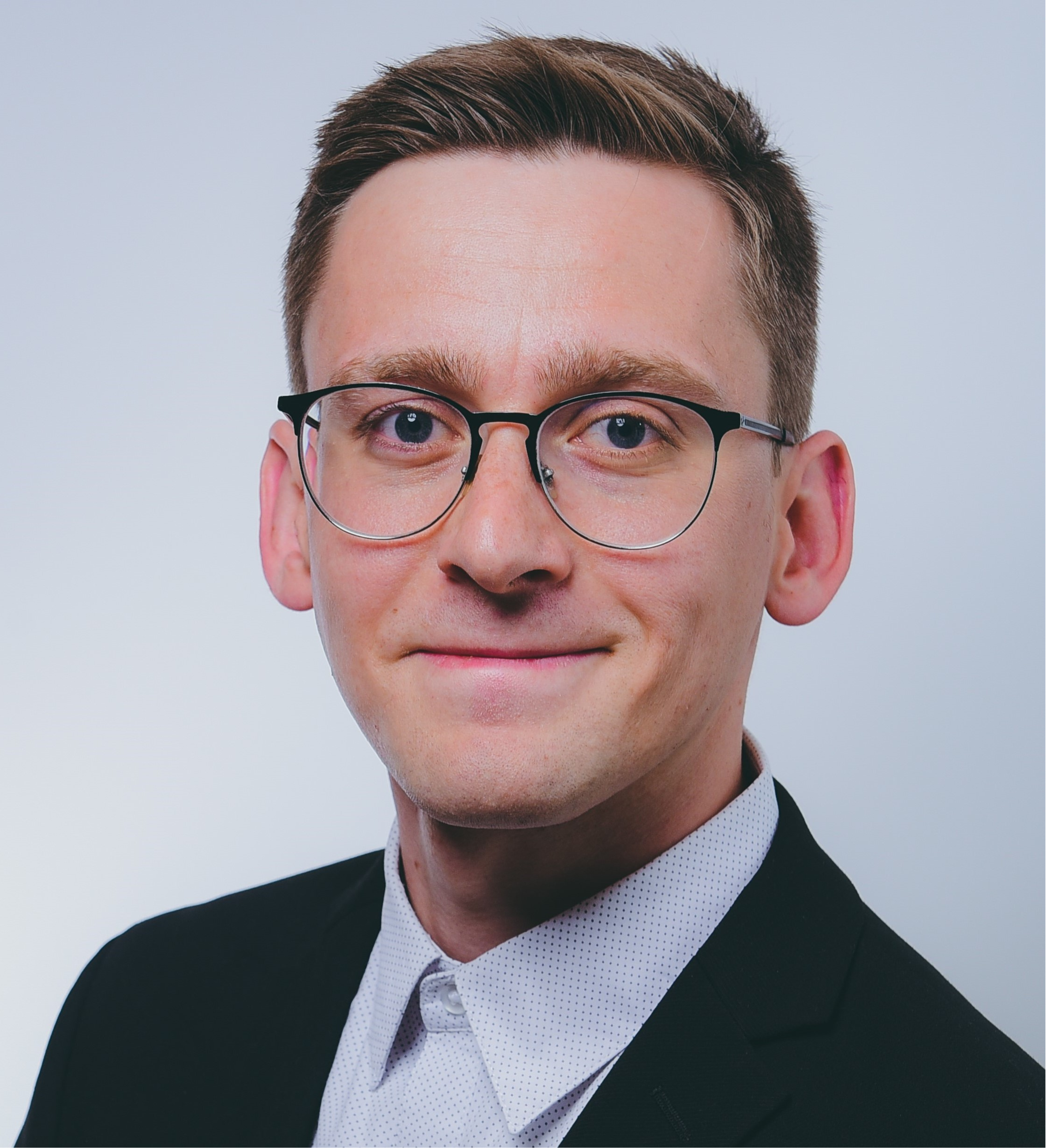}}]{Jakob Holstein}
received first state examination diploma (M.Ed.) for mathematics and physics upper secondary class teaching profession in 2020 from Johann Wolfgang Goethe-University Frankfurt am Main, Germany.
In 2021 and 2022, he received  B.Sc. and M.Sc. degree in physics from the same university.
He is a Ph.D. student in physics in the Ultrafast Spectroscopy and Terahertz Physics Group, Johann Wolfgang Goethe-Universität Frankfurt am Main, Germany. His research interests focus on experimental characterization of Si-CMOS and graphene based TeraFETs for direct power and heterodyne sensing applications. He is working on the realization of integrated TeraFET systems focusing on gas spectroscopy applications.
\end{IEEEbiography}

\begin{IEEEbiography}[{\includegraphics[width=1in,height=1.25in,clip,keepaspectratio]{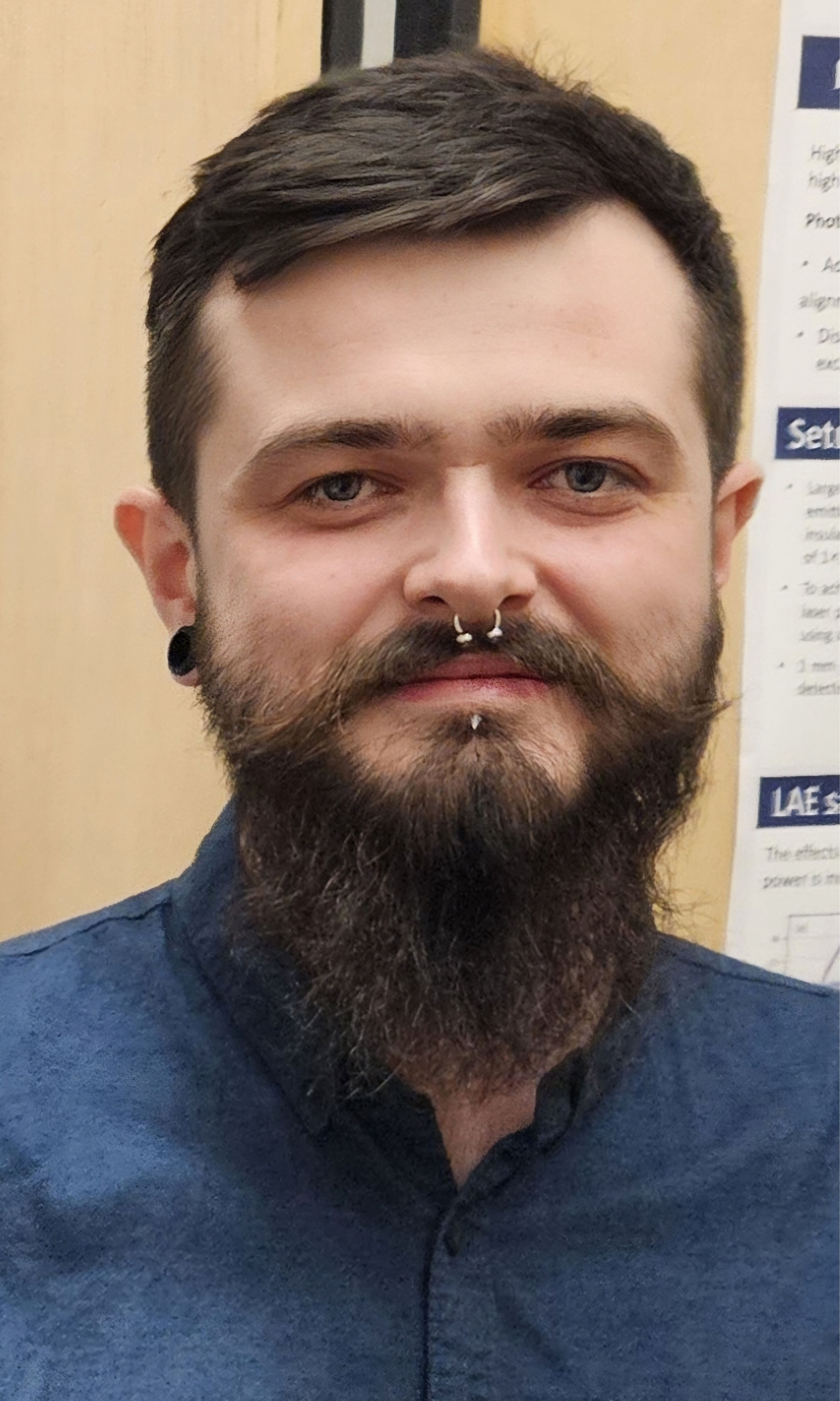}}]{Nicholas North}
received his M.Eng (Hons) degree and PhD degree in electronic and electrical engineering from the University of Leeds, Leeds, U.K., in 2019 and 2024 respectively.

In 2024, he began a research fellow position in the same group. His research focuses on THz high-resolution gas spectroscopy using a QCL as a radiation source and characterising TeraFET detectors with THz QCLs. His research interests are multi-pass optics and sensing atmospheric gas species in the THz spectrum.  He is working on THz gas spectroscopy targeting atmospheric chemical reactions and high-sensitivity gas sensing, where fast and sensitive THz detectors are an interest.
\end{IEEEbiography}

\begin{IEEEbiography}[{\includegraphics[width=1in,height=1.25in,clip,keepaspectratio]{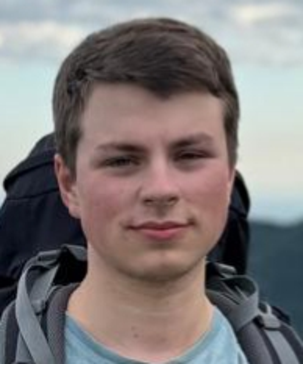}}]{Arne Hof}
received his B.Sc. degree in physics in 2024 from the Johann Wolfgang Goethe-University Frankfurt am Main, Germany.
\end{IEEEbiography}

\begin{IEEEbiography}[{\includegraphics[width=1in,height=1.25in,clip,keepaspectratio]{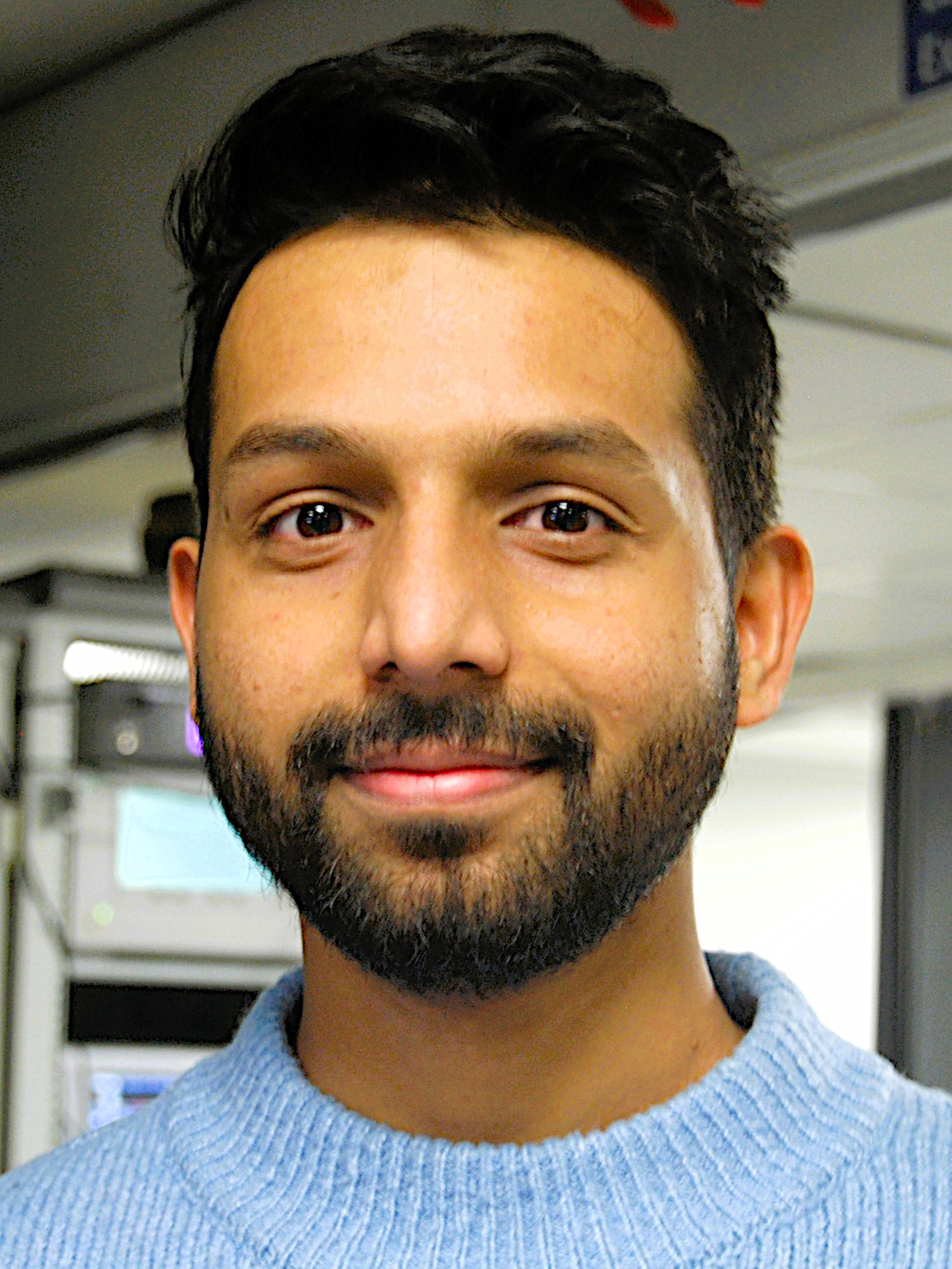}}]{Sanchit Kondawar}
received his B.Tech degree in Electronics and Telecommunication Engineering from the Government College of Engineering, Jalgaon, Maharashtra, India, in 2018. He received his M.Sc(Eng.) and Ph.D. degrees in Electronic and Electrical Engineering from the University of Leeds, United Kingdom, in 2019 and 2024, respectively. Currently, he is a Postdoctoral Research Fellow at the School of Electronic and Electrical Engineering at the University of Leeds. His research interests comprises fabrication and characterisation of quantum cascade lasers for high-speed terahertz imaging and ultra-high capacity wireless communications, coherent detection and manipulation, characterisation of integrated photonic devices, and the development of satellite components for integrated terahertz receiver systems.
\end{IEEEbiography}

\begin{IEEEbiography}[{\includegraphics[width=1in,height=1.25in,clip,keepaspectratio]{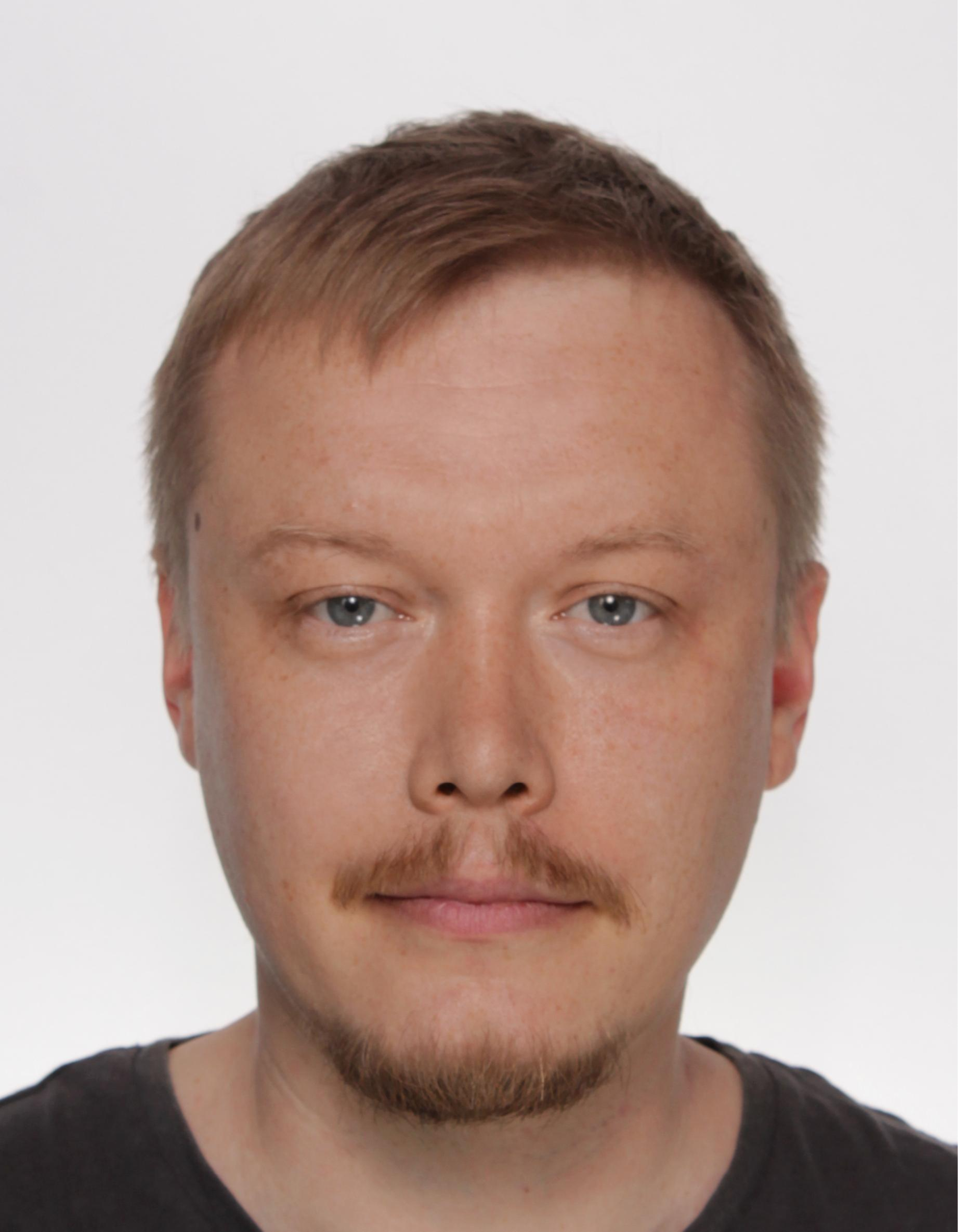}}]{Dymtro D. But}
received the Diploma in microelectronics engineering from the Kyiv Polytechnic Institute NTUU, in Ukraine, in 2009 and received his PhD degree in Physics from the Laboratoire Charles Coulomb, the University of Montpellier II in 2014. In 2014, he also worked as an engineer in the Department of Physics and Technology of Low-dimensional Systems at the V.E. Lashkaryov Institute of Semiconductor Physics NAS of Ukraine. Since 2015 he has continued work as an associate researcher in the group of terahertz spectroscopy at the University of Montpellier. Since 2017, he has been working at the Institute of Electrodynamics, Microwave and Circuit Engineering at the Technical University of Wien as a university assistant. Since 2018, he joined a research group at the Institute of High Pressure Physics of the Polish Academy of Sciences. From 2019, he joined CENTERA Project at the same institution. His research interests include condition matter physics, terahertz electronics, systems and semiconductor devices for them.
\end{IEEEbiography}



\begin{IEEEbiography}[{\includegraphics[width=1in,height=1.25in,clip,keepaspectratio]{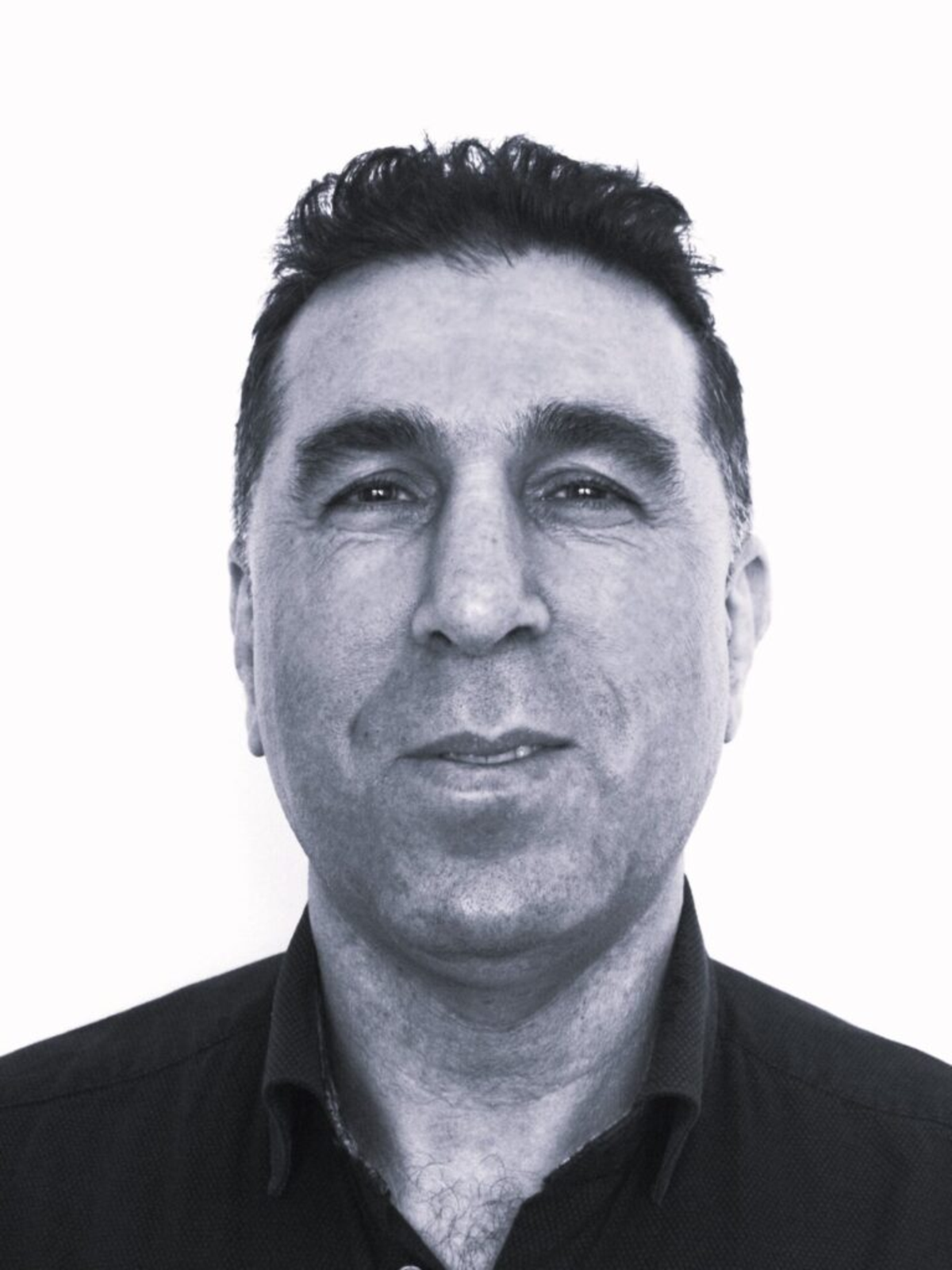}}]{Mohammed Salih}
received the M.Sc. and PhD degree in electronic and electrical engineering from the University of Kassel, Kassel, Germany, in 2002, and from the University of Leeds, Leeds, U.K., in 2011, respectively. He works currently as a Senior  Research Fellow with the University of Leeds, Leeds, U.K. His research interests include the high power density terahertz quantum cascade lasers and utilizing the terahertz wave in space applications.
\end{IEEEbiography}

\begin{IEEEbiography}[{\includegraphics[width=1in,height=1.25in,clip,keepaspectratio]{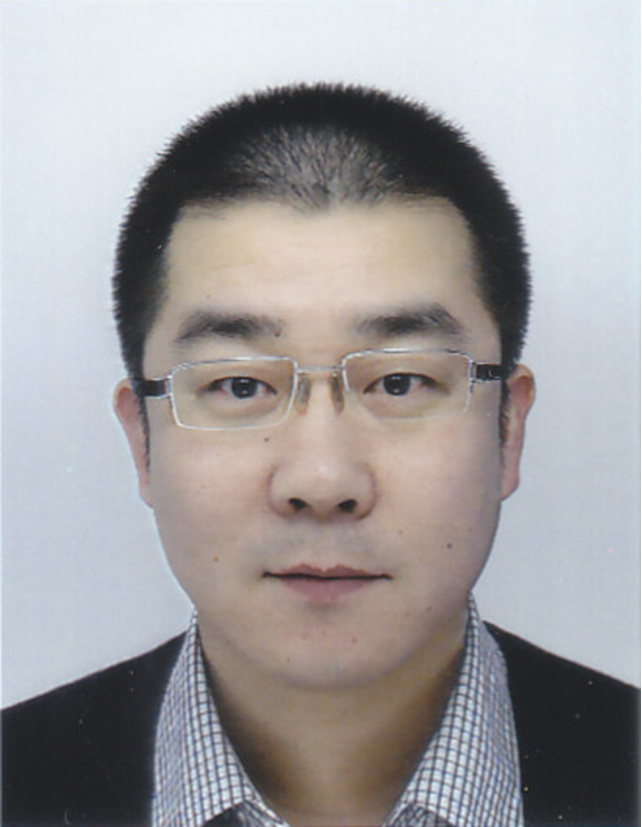}}]{Lianhe Li}
received his Ph.D. degree from Institute of Semiconductors, Chinese Academy of Sciences, in 2001. From 2001 to 2003, he was with the Laboratoire de Photonique et des Nanostructures (CNRS), France, working on the low bandgap III-V diluted nitride materials and devices for telecom applications. In 2003, he joined Institute of Photonics and Quantum Electronics, Ecole Polytechnique Fédérale de Lausanne (EPFL, Switzerland), working on the InAs quantum dots for lasers, superluminescent emitting diodes, and single photon devices. Since 2008, he has been with the school of Electronic and Electrical Engineering, University of Leeds (UK), where his research interests focus on the MBE growth and characterization of semiconductor optoelectronic materials and devices for mid-infrared and THz wave generation/detection with particular emphasis on QCLs, QCDs, QWIPs, ICLs, and LT-GaAs.

\end{IEEEbiography}
\begin{IEEEbiography}[{\includegraphics[width=1in,height=1.25in,clip,keepaspectratio]{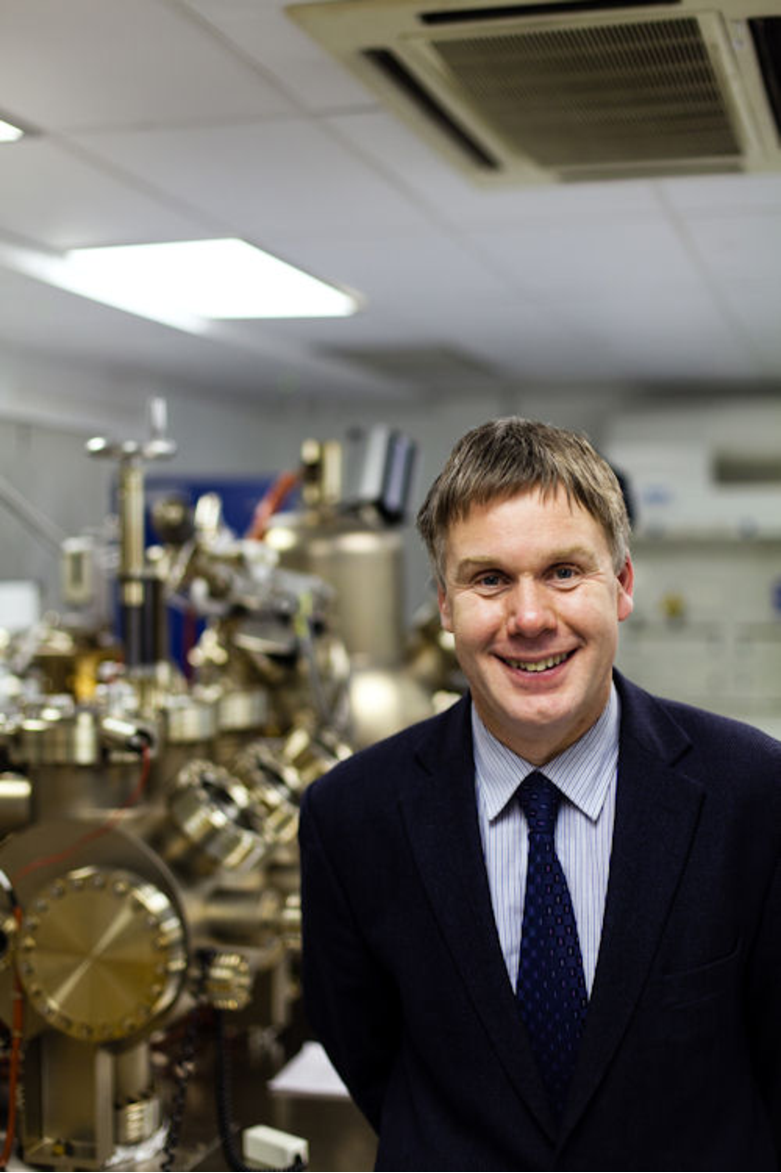}}]{Edmund Linfield}
received the B.A. (Hons.) degree in physics and the Ph.D. degree from the University of Cambridge, Cambridge, U.K., in 1986 and 1991, respectively. He continued his research with the Cavendish Laboratory, University of Cambridge, becoming an Assistant Director of Research and a Fellow of Gonville and Caius College, Cambridge, U.K., in 1997. In 2004, he joined the University of Leeds, Leeds, U.K., to take up the Chair of Terahertz Electronics, and he is currently Director of the University’s ‘Bragg Centre for Materials Research’. He is also ‘Atoms to Devices’ Research Area Lead for the UK’s Royce Institute.  His research interests include semiconductor growth and device fabrication, terahertz-frequency optics and electronics, and nanotechnology.  Prof. Linfield shared the Faraday Medal and Prize from the Institute of Physics in 2014, and was the recipient of the Wolfson Research Merit Award from the Royal Society in 2015.
\end{IEEEbiography}

\begin{IEEEbiography}[{\includegraphics[width=1in,height=1.25in,clip,keepaspectratio]{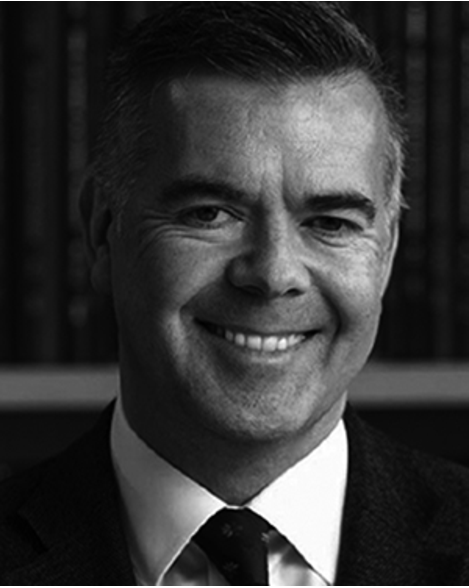}}]{A. Giles Davies}
received the B.Sc. (Hons.) degree in chemical physics from the University of Bristol, Bristol, U.K., in 1987, and the Ph.D. degree in semiconductor physics from the University of Cambridge, Cambridge, U.K., in 1991.
In 1991, he joined the University of New South Wales, Sydney, NSW, Australia, supported by an Australian Research Council Fellowship, before returning to the Cavendish Laboratory, University of Cambridge, in 1995, as a Royal Society University Research Fellow, and subsequently Trevelyan Fellow of Selwyn College, Cambridge. Since 2002, he has been with the School of Electronic and Electrical Engineering, University of Leeds, as Professor of Electronic and Photonic Engineering, and is currently also the Deputy Executive Dean, Faculty of Engineering and Physical Sciences. His research interests include the optical and electronic properties of semiconductor devices, terahertz frequency electronics and photonics, and the exploitation of biological properties for nanostructure engineering.
Professor Davies is a Fellow of the Royal Academy of Engineering and the Institute of Physics, and both a Chartered Physicist and Chartered Engineer. He was the recipient of the Wolfson Research Merit award from the Royal Society in 2011 and shared the Faraday Medal and Prize from the Institute of Physics in 2014.
\end{IEEEbiography}

\begin{IEEEbiography}[{\includegraphics[width=1in,height=1.25in,clip,keepaspectratio]{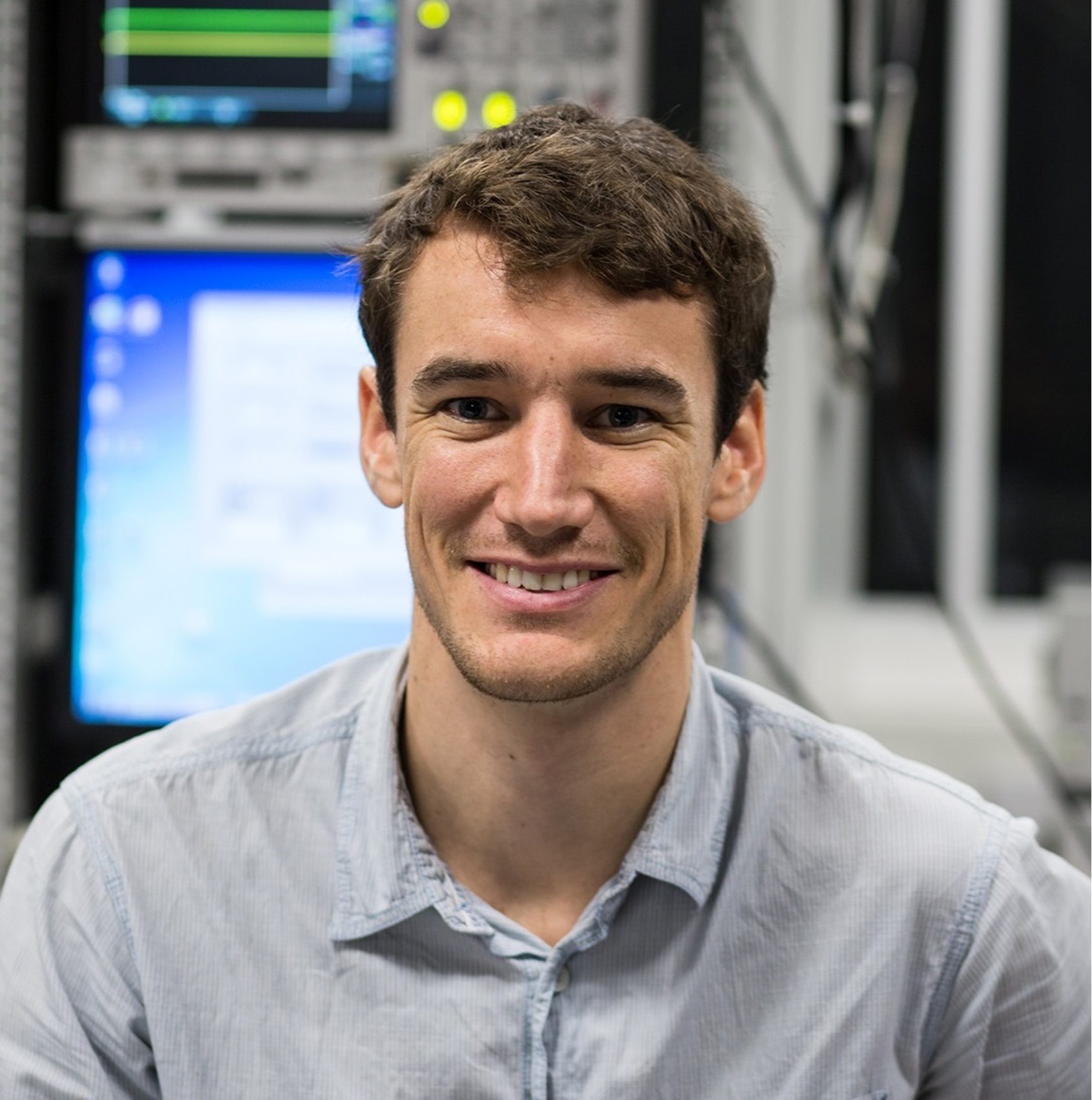}}]{Joshua Freeman}
received the M.Phys. (Hons.) degree in physics from the University Of Oxford, Oxford, U.K., in 2005, and the Ph.D. degree from the University of Cambridge, Cambridge, U.K., in 2010. In 2010, he started a Postdoctoral Fellowship with the Cavendish, supported by the EPSRC. In 2011, won an individual intra-European fellowship and took up the Marie Curie position at the Ecole Normale Supérieure in Paris, returning to the U.K. to take up a postdoctoral research fellowship at the University of Leeds. In April 2016, he was awarded a “University Academic Fellowship” and is currently Associate Professor at the University of Leeds. His research interests include the design, locking, modulation and coherent detection of quantum cascade lasers, and other optoelectronic terahertz devices.
\end{IEEEbiography}

\begin{IEEEbiography}[{\includegraphics[width=1in,height=1.25in,clip,keepaspectratio]{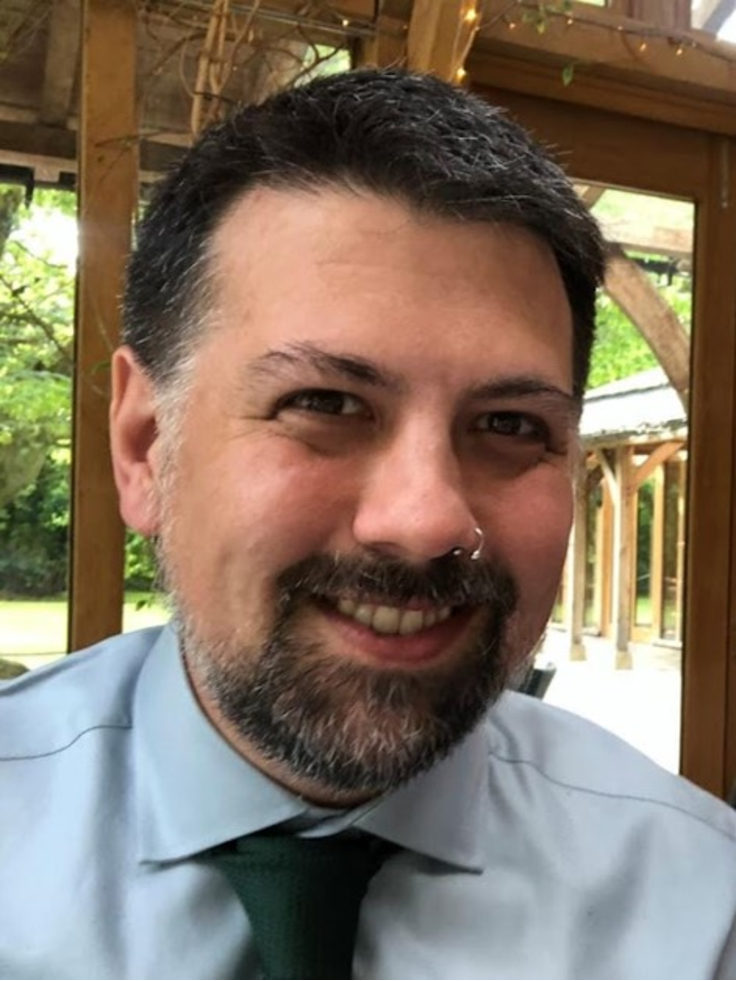}}]{Alexander Valavanis}
received the M.Eng. (Hons) degree in electronic engineering from the University of York, York, U.K., in 2004, and the Ph.D. degree in electronic and electrical engineering from the University of Leeds, Leeds, U.K., in 2009. 

He is an associate professor in terahertz instrumentation at the University of Leeds, Leeds, U.K., and holds a UKRI Future Leader Fellowship.
His research interests include terahertz instrumentation, quantum cascade lasers, remote sensing, gas detection, and computational methods for quantum electronics.

Dr. Valavanis is a member of the Institution of Engineering and Technology (IET). 
\end{IEEEbiography}

\begin{IEEEbiography}[{\includegraphics[width=1in,height=1.25in,clip,keepaspectratio]{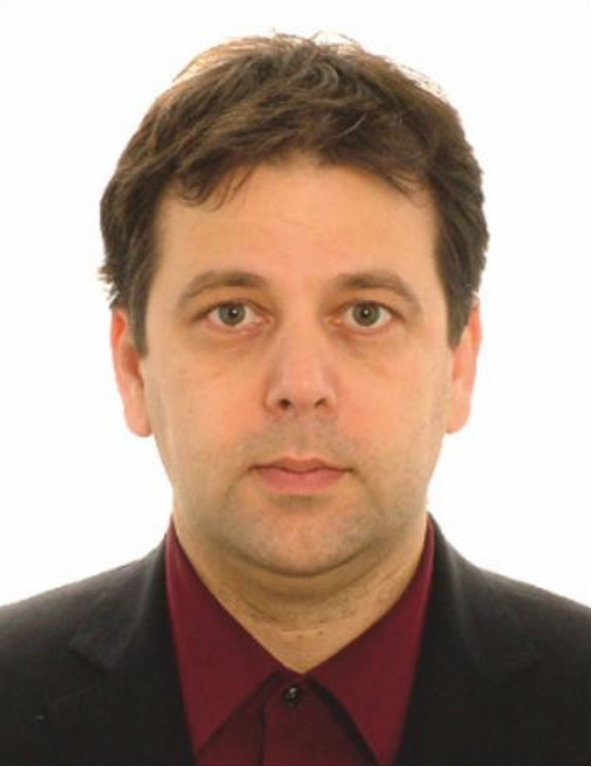}}]{Alvydas Lisauskas}
Prof. IHPP PAS Alvydas Lisauskas received the Diploma in physics from Vilnius University in 1995, and the PhD degree from the Royal Institute of Technology, Stockholm in 2001. In 2002, he joined the Ultrafast Spectroscopy and Terahertz Physics Group at the Goethe University Frankfurt, Germany, working on novel semiconductor devices for THz applications.
From 2014, he is a professor and leading researcher at Vilnius University. From 2014 till 2016 he was a head of joint research laboratory on electrical fluctuations established between the Center for Physical Sciences and Technology and Vilnius University. From 2019 till 2023, he joined CENTERA Project of the Institute of High Pressure Physics PAS, where he led a workgroup on THz electronics.
\end{IEEEbiography}

\begin{IEEEbiography}[{\includegraphics[width=1in,height=1.25in,clip,keepaspectratio]{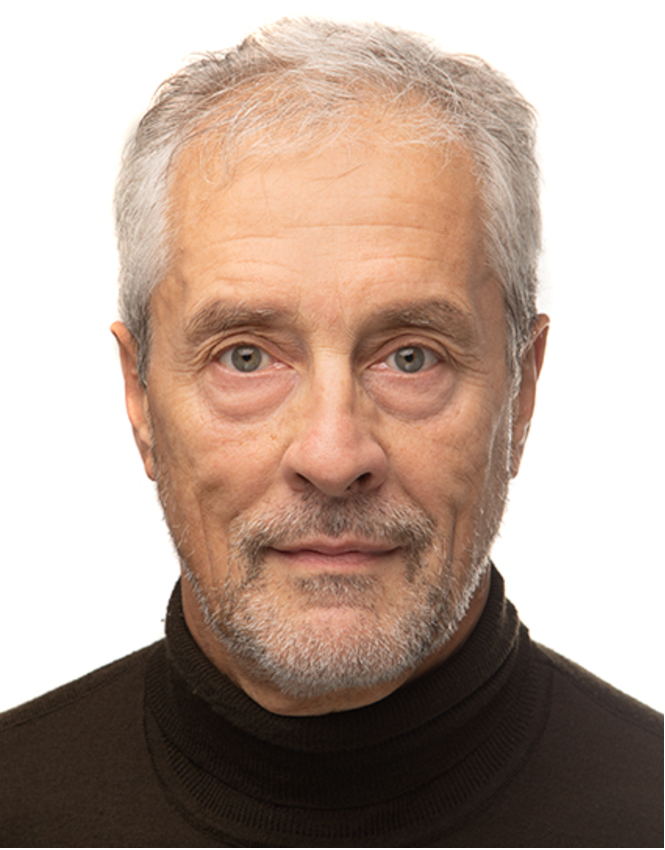}}]{Harmut G. Roskos}
 is Professor of Physics at Johann Wolfgang Goethe-University, Frankfurt am Main, Germany. He studied Physics at the Technical University of Karlsruhe (now Karlsruhe Institute of Technology, KIT, Vordiplom 1981)  and Technical University of Munich (Dipl. phys. degree, 1985), also receiving the PhD degree from the latter in 1989. He then worked at AT\&T Bell Laboratories, Holmdel, USA, where THz phenomena became the focus of his research. He joined the Institute of Semiconductor Electronics of RWTH Aachen in 1991. After receiving the Habilitation degree in Physics from the Faculty of Mathematics and Natural Sciences of RWTH Aachen in 1996 with a thesis on \textit{Coherent Phenomena in Solid-State Physics Investigated by THz Spectroscopy}, he became a Full Professor at Johann Wolfgang Goethe-Universität in 1997. Current fields of work of his group are time-resolved spectroscopy of solid-state materials (mainly materials with strong electronic correlations), strong light-matter coupling with THz metamaterials, metamaterial sensor development, the development of TeraFETs in various device technologies (including graphene FETs), the development of coherently radiating arrays of resonant-tunneling diodes, THz holographic imaging, and the use of Deep Learning for THz imaging.
He spent sabbatical semesters at the University of California at Santa Barbara in 2005 and at the University of Rochester in 2014, and was an Invited Guest Professor at Osaka University’s Institute of Laser Engineering during the winter semester 2009/2010. In 2009, OC Oerlikon AG awarded his group jointly with the Ferdinand-Braun-Institute (FBH) in Berlin a 5-year endowed professorship which led to the establishment of a Joint Lab for THz Photonics.    
\end{IEEEbiography}

\end{document}